\documentclass[aps,pre,twocolumn,floats]{revtex4}
\usepackage{epsfig}
\begin{document}

\newcommand{\tbox}[1]{\mbox{\tiny #1}}
\newcommand{\half}{\mbox{\small $\frac{1}{2}$}}
\newcommand{\sinc}{\mbox{sinc}}
\newcommand{\mbf}[1]{{\mathbf #1}}


\title{Stadium Billiard with Moving Walls}

\author{Doron Cohen$^1$ and Diego A. Wisniacki$^2$}

\date{October 2002}

\affiliation{
$^1$ \mbox{Department of Physics,
Ben-Gurion University,
Beer-Sheva 84105, Israel}\\
$^2$ \mbox{Departamento de F\'{\i}sica "J.J. Giambiagi",
FCEN - UBA, Pabell\'on 1, Ciudad Universitaria,
1428 Buenos Aires, Argentina.}
}


\begin{abstract}
We study the evolution of the energy distribution
for a stadium with moving walls. We consider
one period driving cycle, which is characterized
by an amplitude $A$ and wall velocity $V$.
This evolving energy distribution has
both "parametric" and "stochastic" components.
The latter are important for the theory of quantum
irreversibility and dissipation in driven mesoscopic
devices (eg in the context of quantum computation).
For extremely slow wall velocity $V$ the spreading
mechanism is dominated by transitions
between neighboring levels, while for larger
(non-adiabatic) velocities the spreading
mechanism has both perturbative and non-perturbative features.
We present, for the first time, a numerical study
which is aimed in identifying the latter features.
A procedure is developed for the determination
of the various $V$ regimes. The possible implications
on linear response theory are discussed. \\
\end{abstract}

\maketitle

\section{Introduction}

Consider the problem of particle in a box,
where some piece of the wall is deformed
periodically in time. As an example one
may think of a particle in a cylinder with
a moving piston. The particle has mass $m$
and kinetic energy $E$. The piston is pushed
back and forth. The velocity in which the
piston is displaced is $\pm V$,
and the maximum displacement is $A$.
At the end of each cycle the piston is back
in its original location.

In the present Paper we are going to study
what happens to a quantum mechanical particle
in such a box, during one cycle of the driving.
We assume that the particle is initially prepared
in an energy eigenstate of the (un-deformed) box.
We shall explain that it is important to
specify whether the maximum displacement $A$
is less or larger compared with De-Broglie wavelength.

The problem of particle-in-a-box
with a moving wall is a prototype example
for study of driven systems \cite{lecnotes}
that are described by a Hamiltonian ${\cal H}(x(t))$,
where $x(t)$ is a time dependent parameter.
In case of a piston, the parameter $x$ is
the position of the piston, and we use the
notation $V=|\dot{x}|$.

There is a lot of interest today in
studies of quantum irreversibility.
This is a relevant issue to the design
of quantum computers, where the
fidelity of a driving cycle is important.
If at the end of a driving cycle the system
is back in its initial state, then
we say that the driving cycle has high
fidelity. Obviously, the piston model can
serve as a prototype example for studying
of quantum irreversibility \cite{dsp}.

The study of one-cycle driving also constitutes
a bridge to the study of the response to
multi-cycle (periodic) driving. It should be
clear that the long time behavior of
driven systems is determined by the short time
dynamics. Therefore it is essential to have
a good understanding of the latter.

At first sight one may think that
it is simplest to study the one-dimensional (1D) box case,
also known as `infinite square well potential
with moving wall' \cite{infwell_a,infwell_b}.
The case of periodic driving is also
known as the `Fermi acceleration problem' \cite{jose}.
In a second sight one find out that the 1D-box
case is actually the most complicated one.
As in the case of the kicked rotator (standard map) \cite{qkr}
there is a complicated route to chaos and stochasticity.

Things become much simpler if the motion in the box
is chaotic to begin with. Driven chaotic systems exhibit,
as a result of the driving, stochastic energy spreading
of relatively simple nature \cite{lecnotes}.
This is the case that we want to consider in this Paper.
Hence we consider the simplest "chaotic box", which
is a two dimensional (2D) "billiard" model.
Specifically, we are going to introduce a detailed numerical
study of the {\em one-pulse response of a stadium billiard
to a shape deformation}.
The stadium billiard is a recognized
prototype system for "quantum chaos" studies.
Our simulations are feasible thanks
to a new powerful technique for finding
clusters of billiard eigenstates
\cite{VS,vergini_thesis,barnett_thesis}.
Previous applications of this technique, to the study
of restricted aspects of the present problem,
have been reported in \cite{diego} and in \cite{prm}.

The work which is presented in this Paper,
is the first numerical study which is aimed
in presenting a systematic analysis of non-perturbative
features of a {\em time dependent} spreading
process \cite{crs,rsp,frc}.
A procedure is developed for the determination
of the various time stages in the evolution of the energy
distribution. This allows the identification of the
various $V$ regime ("adiabatic", "perturbative",
"non-perturbative", and "sudden") which were
predicted in past theoretical studies.


\section{Outline}

In {\bf Sections~3} and {\bf Section~4}
we define the model system,
and briefly describe the classical picture.
In particular we explain that the
{\em time dependent features} that we study
in this Paper are purely quantum-mechanical,
and not of semiclassical origin.

In {\bf Section~5} we define the main object of our study,
which is the energy spreading kernel $P_t(n|m)$.
See Eq.(\ref{e3}).
Regarded as a function of the level index $n$,
it is the energy distribution after time $t$,
while $m$ is the initial level.
We also define the sqrt of the variance $\delta E(t)$,
and the $50\%$ probability width $\Gamma(t)$,
that characterize the evolving energy distribution.

Our three phase strategy for analysis of energy spreading
is presented in {\bf Section~6}. The three phases are:
\begin{minipage}{\hsize}
\vspace*{0.2cm}
\begin{itemize}
\setlength{\itemsep}{0cm}
\item[(I)] Study of the band profile.
\item[(II)] Study of parametric evolution.
\item[(III)] Study of the actual time evolution.
\end{itemize}
\vspace*{0.0cm}
\end{minipage}
The relevant information regarding "Phase I"
is summarized in {\bf Section~7}.
Various approximations for $P_t(n|m)$,
and in particular the notion
of "parametric evolution" are
presented in {\bf Section~8}.
The relevant information regarding "Phase II"
is summarized in {\bf Section~9}.
The main concern of this Paper is with "Phase III".

For a given $V$ one should be able to characterize
the nature of the dynamical scenario.
The theoretical considerations regarding this issue
have been discussed in \cite{crs,rsp,frc,vrn},
and for a concise review see \cite{dsp}.
Rather than duplicating these discussions,
we are going to present in {\bf Section~10}
a phenomenological definition, and a practical procedure,
for the identification of the $V$ regimes.
We would like to emphasize that this is the first time
that the different dynamical scenarios (corresponding
to the different $V$ regimes) are illustrated in a numerical
simulation. The only other numerical studies
(mainly \cite{diego,rsp}) were too restricted in scope,
and did not contain analysis of the stages in
the evolution of the energy {\em distribution}.

{\bf Section~11} and {\bf Section~12} question
the applicability of linear response theory (LRT)
to the analysis of the energy spreading.
Further discussion of "non perturbative response",
and conclusions, are presented in {\bf Section~13}.

\section{The semiclassical picture}

Consider a {\em classical particle} inside a box.
Its kinetic energy is $E$ and the corresponding
velocity is  $v_E=\sqrt{2E/m}$.
The shape of the box is externally controlled.
The control parameter is denoted by $x$.
We assume that $x$ has units of length,
such that $V=|\dot{x}|$ is the typical wall
velocity. Obviously different parts of the wall
may have different velocities. In case of the
"piston model" only one piece of wall is moving
(either inward or outward). However, we are
not interested in the trivial {\em conservative} work
which is being done, but only in the {\em irreversible} work.
Therefore, rather than analyzing an actual "piston model"
configuration, it is wiser to consider a {\em volume preserving}
deformation. In such case the conservative work is zero,
and the major issue is the {\em irreversible} effect
which is explained below.

As a results of the collisions of the particle
with the deforming walls, there is a stochastic-like
diffusion in energy space. The explanation is as follows:
Each time that the particle
collides with the moving wall it either
gains or looses energy. To simplify the presentation
let us assume head-on collisions.
The change in energy is $\pm 2mv_EV$ depending on
whether the wall is moving inward or outward
at the point of the collision. For volume
preserving deformation the ergodic average
over $\delta E$ gives zero.  Thus we have random-walk
in energy space where the steps are $\pm 2mv_EV$.
This leads to diffusion in energy space.
As explained in \cite{frc} this diffusion
is biased (the diffusion is stronger for larger $E$).
This leads, in the long run, to a systematic
(irreversible) increase of the average energy.
For a more detailed presentation that takes
the box geometry into account see \cite{wlf}.

Do we have a corresponding (semiclassical) picture
in the quantum-mechanical case?
Again we remind the reader that we assume
a volume preserving deformation. This implies
that the energy levels of the system do not
have a collective "upward" or downward" change
as a result of the deformation.
The physics in which we are interested
is related to the transitions between different
levels. The question that we ask is whether
these transitions are "classical-like".

Let us assume that we start with an eigenstate
whose energy is $E$.
Semiclassically it is as if we start with
a microcanonical preparation. If the dynamics
is of classical nature, then we expect that after
a short time some of the probability will make
a transition $E \mapsto E'$ such that
$|E' - E| \sim 2mv_EV$.  Naturally such description
is meaningful only if the energy scale $2mv_EV$ is
much larger than the mean level spacing.
This leads for 1D-box to the condition
\begin{eqnarray} \label{e1}
V > \hbar/mL
\end{eqnarray}
In the general case (eg 2D box), having
$2mv_EV$ much larger than the mean level spacing
is {\em not} a sufficient condition for getting
semiclassical behavior.
Still, non-trivial analysis \cite{wld}
reveals that the condition for getting semiclassical
spreading in the general case is the same as
in the 1D case, namely given by Eq.(\ref{e1}).

The above Eq.(\ref{e1}) is a necessary condition
for having semiclassical transitions.
In order to actually {\em witness} semiclassical
transitions it is also necessary to have a
time period much larger compared with the ballistic
time ($A/V \gg L/v_E$), and to have
an amplitude that is much larger compared
with De-Broglie wavelength ($A\gg \hbar/(mv_E)$).
These additional conditions can be
satisfied only in the {\em semiclassical regime}
which is defined by Eq.(\ref{e1}),
else they are not compatible.

\section{The numerical model}

As a specific example for chaotic box,
we consider the quarter stadium billiard.
We define $x$ as the length of the straight edge,
and adjust the radius parameter such that
the total area is kept constant.
For numerical reasons we do not analyze this
model "literally" but rather consider,
as in previous study \cite{diego},
a linearized version of the quarter stadium
billiard Hamiltonian.
Namely, we study a model Hamiltonian
that has the matrix representation
\begin{eqnarray} \label{e2}
{\cal H}(x(t)) \mapsto \mbf{E} + \delta x(t) \mbf{F}
\end{eqnarray}
Here $\mbf{E}$ is an ordered diagonal matrix,
that consists of the eigen-energies $E_n$
of the quarter stadium billiard with straight edge $x=1$.
The eigen-energies were determined numerically.
The perturbation due to $\delta x$ deformation,
is represented by the matrix $\mbf{F}$.
Also this matrix has been determined
numerically as explained in \cite{diego}.
We note that the fingerprints of the classical chaos
are present in the statistical properties
of the matrices $\mbf{E}$ (level statistics)
and $\mbf{F}$ (band structure). The latter
is discussed in Section~7.

The {\em linearization} Eq.(\ref{e2})
of the billiard Hamiltonian
can be regarded as a valid {\em approximation} if the wall
displacement parameter $\delta x$ is small compared
with De-Broglie wavelength.
This automatically excludes the possibility
of addressing the semiclassical regime
which has been discussed in the previous section.
In our simulation the De-Broglie wavelength
is roughly $0.1$. This implies that (at best)
the maximum driving amplitude that can be allowed
is $A=0.2$, so as to have $|\delta x(t)| < 0.1$.
[Note that for presentation purpose we later
re-define $\delta x(t)$ as $\delta x(t) - \delta x(0)$.]
In the simulations we indeed have $A=0.2$,
meaning that we allow a deformation which is
comparable or possibly somewhat larger than
that "allowed" by the linearization.

However, inspite of the fact that Eq.(\ref{e2})
may be a "bad" or even "inadequate" approximation
to real-world physics, it is still a totally "legitimate"
Hamiltonian from mathematical point of view.
Moreover, the numerical model Eq.(\ref{e2})
contains all the {\em physical} ingredients that
are relevant for the aim of the present study.

{\em To summarize:} In this paper we consider,
as far as formulation is concerned,
a chaotic billiard driven by volume preserving
shape deformation.
On the other hand, as far as numerics is concerned,
we analyze a specific quantum mechanical
model which is defined by Eq.(\ref{e2}).
The numerical model is motivated by the
quarter stadium billiard system Hamiltonian,
but still it is not literally the
same model. In particular, Eq.(\ref{e2})
does not possess the the semiclassical regime
which has been discussed in the previous section.

\section{The evolving energy distribution}

In case of time independent Hamiltonian,
the energy distribution does not change with time.
In order to have an "evolving" energy distribution,
we have to make $\delta x(t)$ time dependent.
One possibility is to assume linear driving.
In such case we write $\delta x(t)=Vt$.
But more generally, for a cycle,
we write $\delta x(t) = A \times f(t)$,
with the convention $f(0)=f(T)=0$.
In some equations below, whenever a linear driving
is concerned, $A$ can be replaced by $V$,
which assumes the particular choice $f(t)=t$.
For practical purpose, given $f(t)$,
it is convenient to associate with it
the a spectral function
\begin{eqnarray} \label{e_3}
\tilde{F}_t(\omega) \ = \
\left| \int_0^t \dot{f}(t') \mbox{e}^{i\omega t'} dt' \right|^2
\end{eqnarray}
Some useful cases are summarized in Appendix~A.
In this Paper we are primarily interested in the
case of triangular pulse Eq.(\ref{eB6}).

Given the model Hamiltonian Eq.(\ref{e2}), and
the driving scheme $\delta x(t)$, we can calculate
the unitary evolution operator $U(t)$.
This kernel propagates "wavefunctions" in time.
Equivalently, we can describe the quantum
mechanical state using a probability density
matrix. The propagator of the latter
is denoted by $\Lambda(t)$.
In order to describe the evolution
of the energy distribution we define the kernel
\begin{eqnarray} \label{e3}
P_t(n|m)
&=& \mbox{trace}(\rho_n \Lambda(t) \rho_m)
\nonumber \\
&=& |\langle n(x(t)) | U(t) | m(x(0)) \rangle |^2
\end{eqnarray}
where $\rho_n$ can be interpreted as either the probability matrix
or as the corresponding Wigner function that represents the
eigenstate $|n\rangle$. In the latter case the trace operation
should be interpreted as $dPdQ/(2\pi\hbar)^d$ integration
over phase space, where $(Q,P)$ are the canonical coordinates
of the particle, and $d=2$ is the dimensionality.

We shall use the notation $P_t(r)=P_t(n-m)=P_t(n|m)$,
with implicit average over the reference state $m$.
We shall refer to $P_t(r)$ as the "average spreading profile".
Whenever we have a wide distribution, we disregard
the distinction between "energy difference" and
"level difference", and make the identification
\begin{eqnarray} \label{e4}
E_n - E_m  \ \equiv \ \hbar\omega_{nm} \ \approx \ \Delta\times r
\end{eqnarray}
where $\Delta$ is the mean level spacing, and $r=n-m$.
(In our numerics $\Delta \approx 7.22$).

{\bf Fig.1} displays the evolution as a function
of time. Each column is a profile
of the probability distribution $P_t(r)$.
The first~7 time steps are log spaced,
while the rest are linearly spaced.
The $V=100$ evolution is predominantly parametric.
As $V$ becomes smaller and smaller
the deviation from parametric evolution becomes
larger and larger. Some representative
spreading profiles are presented in {\bf Fig.2}.


There are various practical possibilities available
for the characterization of the distribution $P_t(r)$.
It turns out that the major features of this distribution
are captured by the following three measures:
\begin{eqnarray} \label{e5}
{\cal P}(t) & = &  P_t(r=0)
\\ \label{e6}
\Gamma(t) & = & \mbox{50\% probability width}
\\ \label{e7}
\delta E(t) & = & \Delta \times \left(\sum_r r^2 P_t(r)\right)^{1/2}
\end{eqnarray}
The first measure is the survival probability ${\cal P}(t)$.
It is mentioned here just for completeness of presentation.
The second measure $\Gamma(t)$ is the energy width of the
central $r$ region that contains 50\% of the probability.
It is calculated as $\Gamma(t)=(r_{75\%}-r_{25\%})\Delta$,
where $r_{25\%}$ and $r_{75\%}$ are the values
for which the cumulative energy distribution
equals $25\%$ and $75\%$ repectively.
Finally the energy spreading $\delta E(t)$ is
defined above as the square-root of the variance.

{\bf Fig.3} shows how the width $\Gamma(t)$ of the profile evolves.
The lower panel of Fig.3 is a log-log plot.
We see clearly that up to $\delta x_c=0.006$
the width is one level, which is an indication for
the applicability of standard first-order perturbation
theory. For larger $\delta x$ several levels are mixed
non-perturbatively, and therefore the width becomes
larger than one.

{\bf Fig.4} shows how the spreading $\delta E(t)$
of the profile evolves.
The lower panel displays the relative spreading,
which is defined as the ratio between the actual
spreading and the parametric one.
The notion of "parametric evolution" is defined
in the next paragraph.
As $V$ becomes smaller, the departure from
the parametric behavior happens earlier.


The "sudden limit" ($V\rightarrow\infty$)
of $P_t(n|m)$ will be denoted by $P(n|m)$.
We shall refer to it as the
"parametric kernel". It is formally obtained
by the replacement $\Lambda(t)\mapsto 1$,
or equivalently $U(t)\mapsto 1$ in Eq.(\ref{e3}), namely,
\begin{eqnarray} \label{e3p}
P(n|m)
&=& \mbox{trace}(\rho_n\rho_m)
\nonumber \\
&=& |\langle n(x(t)) | m(x(0)) \rangle |^2
\end{eqnarray}
Obviously its dependence on $t$ is exclusively
via $\delta x = x(t)-x(0)$, irrespective of $V$.
The parametric evolution of $P(r)$ versus $\delta x$
for a deforming billiard has been studied in \cite{prm}.
The numerics in this Paper can be regarded
as the extension of the numerics of \cite{prm}
to the case of finite~$V$.

\section{Three phase strategy}

In the future we want to have a theory that allows
the prediction of the (numerically simulated)
time evolution. The minimum input which
is required for such theory is the band profile
of the $\mbf{F}$ matrix.
For precise definition of the bandprofile see Appendix B.
Note the existence of a very efficient
semiclassical recipe to find the band profile.
This can save the need for a tedious quantum mechanical
calculations.

Using the bandprofile one hopes to be able,
in the "second phase" \cite{lds},
to calculated the parametric kernel $P(n|m)$.
The bandprofile does not
contain information about the correlations
between the off-diagonal elements. Therefore
one has to make the so-called RMT conjecture,
namely  to assume that the off-diagonal elements
are effectively uncorrelated.
It turns out that, upon using such conjecture,
the non-universal features of the parametric
evolution are lost.
Still, the obtained results are qualitatively correct,
and therefore such an approach is legitimate as
an approximation.

However, finding the parametric kernel $P(n|m)$,
using the bandprofile as an input, is not the subject
of this paper. Therefore, as a matter of strategy,
we would like to take the numerically determined
parametric evolution as an input.
Given the parametric $P(n|m)$, the question
is whether we can calculate the
actual evolution $P_t(n|m)$ for any finite $V$.

In previous works (see review in \cite{dsp})
we gave a negative answer to the above question.
We have claimed that non-perturbative features
of the dynamics cannot be deduced from the parametric analysis.
To explain this observation let us assume
that we have two model Hamiltonians,
say ${\cal H}_{\tbox{physical}}(x)$
and ${\cal H}_{\tbox{artificial}}(x)$.
Let us assume further that the two models have
the same bandprofile and the same parametric kernel $P(n|m)$,
Still we claim that for finite $V$ the two Hamiltonians may
generate {\em different} temporal kernels $P_t(n|m)$.
In particular, it has been argued that
this is in fact the case if ${\cal H}_{\tbox{artificial}}(x)$
is an effective RMT model which is associated
with  ${\cal H}_{\tbox{physical}}(x)$.

In past publications the manifestation of non-perturbative
features in case of driven physical (non-RMT) models
has not been investigated numerically.
On the theoretical side it is an open
subject for further research \cite{rsp}.
The purpose of the following sections is to present
a strategy for analysis of numerical simulations,
that paves the way towards a theory
for the non-perturbative aspects of the energy spreading.

\section{The bandprofile}

The bandprofile of $\mbf{F}$ is described
by the spectral function $\tilde{C}(\omega)$,
which is a Fourier transform of a correlation
function $C(\tau)$. See Appendix B.
If the collisions are uncorrelated, as in the
case where the deformation involves only a small
surface element, then $C(\tau)$ is a delta function,
and the band profile is flat.
This is not the case in the present model.
The correlations between collisions cannot
be neglected, and therefore $C(\tau)$ equals
delta function plus a smooth component \cite{wlf}.
Due to the smooth component the band profile
has a very pronounced non-universal structure.
See Fig.5.

For large enough $\omega$ the contribution
of the non-universal component vanishes.
Thus the bandprofile should possess flat tails.
These tails reflect the presence of the
delta-function component in $C(\tau)$.
For numerical reasons, due to truncation,
the bandprofile of our system is multiplied
by a Gaussian envelope. Therefore the tails of
the effective bandprofile are not flat,
but rather vanishingly small.
The lack of flat tails in the numerical
model can be loosely interpreted as
having 'soft' rather than 'hard' walls.

Thus our system is characterized by
a finite bandwidth. This is actually the generic case.
Namely, for generic ("smooth") Hamiltonian
the correlation function $C(\tau)$ is non-singular,
which implies finite bandwidth.

\section{Approximations}

A major issue in the studies of energy spreading
is the knowledge how to combine tools or approximations
in order to understand or calculate the kernel $P_t(n|m)$.
In particular we have the following approximations:
\begin{minipage}{\hsize}
\vspace*{0.2cm}
\begin{itemize}
\setlength{\itemsep}{0cm}
\item The perturbative kernel $P_t^{\tbox{prt}}(n|m)$.
\item The semiclassical kernel $P_t^{\tbox{scl}}(n|m)$.
\item The Gaussian kernel $P_t^{\tbox{sto}}(n|m)$.
\item The parametric kernel $P(n|m)$.
\end{itemize}
\vspace*{0.0cm}
\end{minipage}
The parametric kernel $P(n|m)$, that corresponds
to the sudden limit ($V\rightarrow\infty$) has already
been defined in Eq.(\ref{e3p}). The other kernels
will be defined in the present section.
Obviously the kernels listed above are very different.
Our aim is to clarify in what regime ($V$),
in what time stage ($t$), and in what energy region ($r$),
which of them is the valid approximation.

The semiclassical kernel $P_t^{\tbox{scl}}(n|m)$
is defined and obtained by assuming in Eq.(\ref{e3})
that the Wigner functions can be approximated by
smeared microcanonical distributions.
See Ref.\cite{vrn,lds,prm} for further details
and numerical examples.
Whenever  $P_t(n|m)\sim P_t^{\tbox{scl}}(n|m)$
we say that the spreading profile is "semiclassical".
In order to have a valid semiclassical approximation
in case of billiard systems the displacement $\delta x$
of the walls should be much larger than
De-Broglie wavelength \cite{prm}.
Thus, for reasons that were explained in Section~3,
the semiclassical approximation is not applicable
for the numerical model that we consider in this Paper.

The perturbative kernel $P_t^{\tbox{prt}}(n|m)$
is obtained from perturbation theory \cite{frc,vrn}.
The defining expression is:
\begin{eqnarray} \label{e8}
P_t^{\tbox{prt}}(n|m) =&
A^2 \times
\tilde{F}_t\left(\frac{E_n{-}E_m}{\hbar}\right)
\times \nonumber \\ \ &
\frac{\Delta}{2\pi\hbar}
\tilde{C}\left(\frac{E_n{-}E_m}{\hbar}\right)
\ \frac{1}{\Gamma^2 + (E_n{-}E_m)^2}
\end{eqnarray}
where $\Gamma$ is determined by {\em normalization}.
If we make the replacement $\Gamma \mapsto 0$
we get the standard result of first order
perturbation theory (see Eq.(\ref{eD5})).
The presence of $\Gamma$ in the denominator reflects
corrections to infinite order.
We note that $P_t^{\tbox{prt}}(n|m)$ constitutes
a generalization of Wigner Lorentzian.
We indeed would get from it a Lorentzian
if the bandprofile were flat with no finite bandwidth.

Whenever $P_t(n|m)\sim P_t^{\tbox{prt}}(n|m)$
we say that the spreading profile is "perturbative".
The perturbative structure is characterized by having
separation of scales:
\begin{eqnarray} \label{e9}
\Gamma(t) \ \ \ll \ \ \delta E(t)
\end{eqnarray}
We say that $P_t(n|m)$ has a "standard" perturbative
structure Eq.(\ref{eD5}), which is given by first order
perturbation theory, if $\Gamma<\Delta$.
This means that more than $50\%$ probability
is concentrated in the initial level.
We use the term "core-tail" structure if we
want to emphasize the existence of a finite
non-perturbative "core" region $|r| < \Gamma/\Delta$.
The core width $\Gamma$,
as determined by "perturbation theory"
by imposing normalization on Eq.(\ref{e8}),
constitutes a rough estimate for the width $\Gamma(t)$
of the energy distribution (Eq.(\ref{e6})).

The spreading $\delta E(t)$ for the core-tail
structure is determined by the tail region,
which is defined as $|r| \gg \Gamma/\Delta$.
If Eq.(\ref{e8}) were a Lorentzian, it would
imply $\delta E(t)=\infty$.  This is of course
not the case, because the spectral functions
provide a physical cutoff.
Thus, the core-tail structure which is described
by Eq.(\ref{e8}) is characterized by a "tail" component
that contains a vanishingly small probability but still
dominates the variance.

If we do not have the separation of energy scales Eq.(\ref{e9}),
then perturbation theory becomes useless \cite{vrn,frc}.
In such case $P_t(n|m)$ becomes purely non-perturbative.
In order to determine $P_t(n|m)$ we have to use tools
that go beyond perturbation theory. In particular,
for long times one can justify \cite{vrn,frc}
a stochastic approximation, leading to the Gaussian kernel
\begin{eqnarray} \label{e10}
P_t^{\tbox{sto}}(n|m) \ = \
\frac{1}{\sqrt{2\pi}}
\frac{\Delta}{\delta E(t)}
\exp\left[
-\frac{1}{2}\left(\frac{E_n-E_m}{\delta E(t)}\right)^2
\right]
\end{eqnarray}
Note that for very long times this Gaussian
should be replaced by an appropriate solution
of a diffusion equation. But this is not
relevant to our simulations. For a critical discussion
that incorporates estimates for the relevant time
scales see \cite{vrn,frc}.

A final remark: For a Gaussian profile
the $50\%$~width is $\Gamma(t) = 1.35 \times \delta E(t)$.
However, whenever a non-perturbative structure
is concerned  it is better, in order to avoid confusion,
not to use the notation $\Gamma(t)$.
The notation $\Gamma$ has been adopted in the common
diagrammatic formulation of perturbation theory.
In case of non-perturbative structure this formulation
becomes useless, and therefore the significance
of $\Gamma$, as a distinct energy scale, is lost.

\section{Analysis of the Parametric evolution, $\delta x$~regimes}

The parametric evolution of $P(n|m)$ is illustrated
in Fig.1 (upper panel).
For very small $\delta x$ we observe clearly
a standard perturbative profile (Eq.(\ref{eD5})),
whose structure is just a reflection
of the bandprofile. By definition this
holds in the standard perturbative (parametric) regime
$\delta x < \delta x_c$.
From the numerics
(looking on the lower panel of either Fig.3 or Fig.6)
we find that $\delta x_c = 0.006$.

For $\delta x > \delta x_c$ the initial
level starts to mix with neighboring levels.
As a result a non-perturbative `core` component
starts to develop. Thus we obtain a core-tail
structure. The tail is the perturbative
component. The main component of the tail is
the "first order tail" which can be calculated
using first order perturbation theory.
The tails in the vicinity of the core are growing slower,
which can be regarded as a suppression of core-to-tail
transitions due to the mixing.
There are also higher order tails \cite{lds}
that can be neglected.

%
Having  $\delta E \gg \Gamma$ is an indication
that $\delta E$ is dominated by the (perturbative) tail
component of the core-tail structure Eq.(\ref{e8}).
The condition $\Gamma \ll \delta E$,
can be re-written as $\delta x  \ll \delta x_{\tbox{prt}}$,
which constitutes a definition of
the extended perturbative (parametric) regime.
Unfortunately, the strong inequality $\Gamma \ll \delta E$
is nowhere satisfied in our numerics. Therefore
the numerical definition of $\delta x_{\tbox{prt}}$
becomes ambiguous. One may naively define
$\delta x_{\tbox{prt}}$ by transforming the
weak inequality $\Gamma < \delta E$ into
a weak inequality $\delta x  < \delta x_{\tbox{prt}}$.
But this procedure is numerically meaningless:
The quantities $\Gamma$ and $\delta E$ are
energy scales. As such their definition is
arbitrary up to a prefactor of order unity.
After some thinking one realizes that the only
practical definition for $\delta x_{\tbox{prt}}$
is as the $\delta x$ where the $P_t^{\tbox{prt}}$ based
prediction of $\delta E$
becomes significantly less than the actual spreading.
From the numerics (Fig.6) we find
that $\delta x_{\tbox{prt}}=0.05$.
This definition is not ambiguous numerically
because we compare "variance based measure" ($\delta E$)
to "variance based measure" ($\delta E^{\tbox{prt}}$),
rather than comparing "variance based measure" ($\delta E$)
to a "width measure" ($\Gamma$).

%
For $\delta x \gg \delta x_{\tbox{prt}}$
the $P_t^{\tbox{prt}}$ based calculation
of $\delta E$ gives saturation. This (non-physical)
saturation is the consequence of having finite
bandwidth. It should be regarded as an artifact
that reflects the limited validity of perturbation theory.

%
In the non-perturbative (parametric) regime,
namely for  $\delta x > \delta x_{\tbox{prt}}$,
the tails are no longer the dominant
component.  This is associated with a structural
change in the spreading profile.
Looking at the upper panel of Fig.1
we observe a core-tail structure
up to the 13th time step. Then, the secondary lobe
of the bandprofile is swallowed by the core.
This happening is reflected in Fig.6 (upper panel)
by a boost in the width $\Gamma$.
[The notion of "secondary lobe" should be
clear by looking at the $|r|\sim 20$ region
in the upper panel of Fig.2.
Unfortunately its visibility in the
corresponding image (upper panel of Fig.1)
is quite poor.
Still it is possible to follow the evolution
of the $r\sim 20$ tail region, and to
realize that it is swallowed by the core].

%
To have a weak inequality
$\delta x > \delta x_{\tbox{prt}}$
is not quite the same as to be
in the (deep) non perturbative regime
where $\delta x \gg \delta x_{\tbox{prt}}$.
For $\delta x > \delta x_{\tbox{prt}}$
we do not get a purely non-perturbative structure.
Inspite of the failure of the core-tail
picture Eq.(\ref{e8}),
we still can make a {\em phenomenological}
distinction between "core" and "tail" regions.
One clearly observes that the tail region
is "pushed" outside because of the expanding core.
This non-perturbative effect is not captured by Eq.(\ref{e8}).
In a previous study \cite{prm} the crossover
from a core-tail structure to a purely
non-perturbative (semiclassical) structure
was quite abrupt. In the present study
the "deep" non-perturbative regime,
where all the tail components disappear,
is not accessible due to numerical limitations.
A "purely non perturbative"
structure would be obtained if all the tail
components were swallowed by the expanding core.

\section{Analysis of actual evolution, $V$~regimes}

We start this section with a qualitative
description of the actual evolution,
which is based on looking in illustrations
such as in Fig.1 and Fig.2. Later we present
the quantitative analysis.
In general one observes that for finite~$V$
the evolution acquires stochastic-like features.
At intermediate times ($0<t<T$) the spreading
profile contains both parametric and stochastic "components".
The parametric component is "reversible". In contrast
to that the stochastic component of the spreading
is not affected by the velocity reversal, and may
lead to a final Gaussian distribution.

Let us describe what happens to the evolution,
as a function of $\delta x=Vt$,
as we make simulations with smaller and smaller~$V$.
On the basis of Eq.(\ref{e8}) we
expect to observe a modulation of the
tails of $P_t(r)$ in a way that is implied
by the presence of envelope $\tilde{F}_t(\omega)$.
This modulation is characterized
by the energy scale $\hbar/t = \hbar V /\delta x$.
If we make $V$ further and further smaller,
the secondary lobes of the tails
[see previous section for definition]
do not have a chance to become visible,
because they are suppressed
by the narrower envelope $\tilde{F}_t(\omega)$.
For small enough $V$ only the main core
component of the bandprofile is left visible.
For very small $V$ the spreading
profile becomes very close to a Gaussian shape
in a very early stage of the evolution.


The spreading at the end of the pulse period
is very small both is the sudden limit
(very large $V$, corresponding to
multi-period driving with large frequency),
and also in the adiabatic limit (very small $V$).
Fig.2 illustrates representative spreading profiles,
which are observed in the end of the pulse period.
Also displayed are the first-order perturbative
profile (Eq.(\ref{eD5})),
and a Gaussian with the same width. Note that
the first-order perturbative profile Eq.(\ref{eD5}),
unlike Eq.(\ref{e8}), does not have a proper normalization.
We can define a `difference measure' in order
to quantify the deviation of the (actual) lineshape
from Gaussian shape. One possible definition is
\begin{eqnarray} \label{e_diff}
\sqrt{ \sum_r P^{\tbox{sto}}(r) \ [\log(P_t(r))-\log(P_t^{\tbox{sto}}(r))]^2 }
\end{eqnarray}
The lower panel of Fig.8 shows
that for $V<20$ the spreading profile
at the end of the pulse is close to Gaussian shape,
while for $V>20$ this profile is predominantly
of perturbative nature.


We should address now the issue of $V$ regimes.
Namely, for a given $V$ we would like to
characterize the nature of the dynamical scenario.
Below we shall introduce a practical procedure
for the identification of the various regimes.
We are going to explain that in the numerical simulation
we observe four different $V$ regimes:
\begin{minipage}{\hsize}
\vspace*{0.2cm}
\begin{itemize}
\setlength{\itemsep}{0cm}
\item Adiabatic regime ($V<3$).
\item Perturbative regime ($3<V<7$).
\item Non-perturbative regime ($7<V<20$).
\item Sudden regime ($20<V$).
\end{itemize}
\vspace*{0.0cm}
\end{minipage}
Before we get into details, we would like to make
a connection with the theoretical discussion of
regimes in \cite{rsp}.
There we have considered a (sinusoidal) periodic
driving that is characterized by amplitude $A$
and frequency $\Omega=2\pi/T$.
For sinusoidal driving the root-mean square
"velocity" is $V = A\Omega/\sqrt{2}$.
So fixing $A$ and changing $V$ in the present
paper, is completely analogous to fixing $A$
and changing $\Omega$ in Ref.\cite{rsp}.
Thus, the four $V$ regimes listed above correspond
to a horizontal cut ($A=\mbox{const}>A_{\tbox{prt}}$)
in the $(\Omega,A)$ regime diagram
of \cite{rsp} (see Fig.~1 there).
One should realize that the "perturbative regime"
is in fact the "linear response (Kubo) regime"
as defined in \cite{rsp},
and that the "sudden regime" corresponds to the
regime of vanishing (high frequency) response.

In the adiabatic $V$ regime the spreading is due
to near neighbor level transitions, and disregarding
extremely short times (for which we can apply
standard first order perturbation theory (Eq.(\ref{eD5})),
it looks stochastic. This means that Eq.(\ref{e10})
is a quite satisfying approximation. For further
details regarding the identification of the
adiabatic regime see Section~12.

Outside of the adiabatic regime there is a time
stage where perturbation theory (Eq.(\ref{e8})) is valid.
But after this time we have to use
theoretical considerations that go
beyond perturbation theory. In the following
paragraph we divide the non-adiabatic regime into
"perturbative" $V$~regime,
"non-perturbative" $V$~regime,
and "sudden" regime.
The basis for this distinction is the
timing of the {\em departure} from parametric evolution.
The departure from parametric evolution
is best illustrated by the lower panel of Fig.4.

The perturbative $V$ regime is defined by the
requirement of having the {\em departure}
from parametric evolution happen {\em before}
the breakdown of Eq.(\ref{e8}).
Consequently, in this regime the departure
time can be deduced from Eq.(\ref{e8}).
For example, let us assume a simple bandprofile.
In such case departure time is just the inverse
of the bandwidth, which implies via Eq.(\ref{eA3})
that it is simply the classical correlation
time $\tau_{cl}$ of the chaotic motion.
In the actual numerical analysis the bandprofile
is not "simple", but has some structure.
Rather than debating over the definition of $\tau_{cl}$,
it is more practical to determine the departure time
by inspection of Fig.4.
For convenience we have indicated the location
of $\delta x_{\tbox{prt}}$ by a vertical line.
For $V<7$ the departure from parametric evolution
happens before this line. This way one can determine
the upper border ($V=7$) of the perturbative regime.

The non-perturbative $V$ regime is defined by the
requirement of having the {\em departure}
from parametric evolution happen {\em after}
the breakdown of Eq.(\ref{e8}).
In other words, it means that this departure cannot
be captured by perturbation theory.
A more careful definition of the non-perturbative
regime should take into account the driving reversal
at $t=T/2$. As explained in the next paragraph,
the non-perturbative regime is further restricted
by the requirement of having the {\em departure}
from parametric evolution happens {\em before}
the driving reversal.

The distinction between "perturbative" and
"non-perturbative" $V$~regimes is related
to the possibility to "capture" the {\em stochastic}
features of the energy spreading by perturbation
theory. Non-perturbative features of the
energy distribution during intermediate times
are of no relevance if they are of parametric
(non-stochastic) origin. Let us look for example,
in the upper panel of Fig.1.
We clearly see that the spreading
profile at the end of the pulse (at $t=T$)
is of standard perturbative nature.
All the non-perturbative features
that develop during the first half
of the driving cycle, are completely
reversed in the second half of the cycle.

Conceptually the simplest way to set a criterion for
getting at the end of the pulse a perturbative structure,
is to use "fixed basis" perturbation theory
(see Appendix~C of \cite{frc}).
If we adhere to the present (more physical) approach
of using $x$-dependent basis, an equivalent method \cite{frc}
is to determine the sudden time $t_{\tbox{sdn}}$.
This is the time to resolve the expanding "core".
It is determined as the time when the following inequality
is violated:
\begin{eqnarray} \label{e11}
\Gamma(t) \ \ \ll \ \ \hbar/t
\end{eqnarray}

The regime where the condition $t_{\tbox{sdn}} \ll T$
is violated is defined as the "sudden regime".
In the sudden regime all the non-perturbative features
are of parametric nature, and therefore the validity
of perturbation theory survives at the end of the pulse.
In the sudden regime the {\em departure}
from parametric evolution becomes
visible only {\em after} the driving reversal.
One can determine the sudden regime
simply by looking at the lower panel of Fig.8.
The departure of the energy distribution (at $t=T$)
from Gaussian shape is correlated
with getting into the sudden regime.

The discussion above of $V$~regimes was based on looking
for the {\em breaktime} of perturbation theory Eq.(\ref{e8}),
on the one hand, and looking for the {\em departure}
from parametric behavior on the other hand.
The departure from parametric behavior is an indication
for the appearance of a predominant stochastic component
in the spreading profile. This departure does not imply
that we get a Gaussian line shape.
In the perturbative regime we get the Gaussian line
shape {\em after} the breaktime of perturbation theory,
which happens {\em after} the departure from parametric
behavior. Therefore in the perturbative regime
there are 3~stages in the evolution:
a parametric stage, a perturbative stochastic stage,
and a genuine stochastic stage.
In contrast to that, in the non-perturbative regime
we do not have an intermediate "perturbative stochastic stage",
because the departure form parametric behavior
happens {\em after} the breakdown of perturbation theory.

\section{LRT formula}

The general LRT formula for the variance of
the spreading is
\begin{eqnarray} \label{e12}
\delta E(t)^2 \ = \ A^2 \times
\int_{-\infty}^{\infty} \frac{d\omega}{2\pi}
\tilde{F}_t(\omega) \tilde{C}(\omega)
\end{eqnarray}
The proportionality  $\delta E \propto A$
reflects having "linear response".
Two spectral functions are involved:
One is the the spectral content
of the driving (Appendix~A),
and the other is the power spectrum
of the fluctuations (Appendix~B).
The latter is the Fourier transform
of a correlation function $C(\tau)$.

A special case is the sudden limit
($V=\infty$) for which $\tilde{F}_t(\omega)=1$
and accordingly
\begin{eqnarray} \label{e13}
\delta E(t) \ = \ \sqrt{C(\tau=0)} \times A
\end{eqnarray}
Another special case is the response for
persistent (either linear or periodic) driving
where $\tilde{F}_t(\omega) = t \times 2\pi\delta(\omega)$
implies diffusive behavior:
\begin{eqnarray} \label{e14}
\delta E = \sqrt{2 D_E t}
\end{eqnarray}
In such case the expression for $D_E$ is known
as Kubo formula, leading to a fluctuation-dissipation
relation.

Finding the conditions for validity of LRT is
of major importance. This should be regarded as
the first step in the analysis of the response
of a driven system. The classical derivation of
Eq.(\ref{e12}) is quite simple, and for completeness we
present it in Appendix~C. For sake of the following
discussion one can assume that the classical "slowness"
conditions for the validity of classical LRT are satisfied.

The quantum mechanical derivation of this formula
is much more subtle \cite{crs,frc,vrn,rsp} and leads
to the distinction between "adiabatic"
and "(extended) perturbative"
and "non-perturbative" regimes.
The semiclassical limit is contained in the latter
regime. It is known that the LRT formula
does not hold in the adiabatic regime \cite{wilk},
but it is valid in the (extended) perturbative regime \cite{crs,frc}.
It is not necessarily valid in the non-perturbative
regime \cite{crs,rsp}, but if the system has a classical limit,
it must be valid again in the semiclassical regime.
See further discussion in the next sections.

\section{Analysis: Comparison with LRT}

Using the bandprofile as an input we
can calculate the spreading using Eq.(\ref{e12}).
In Fig.6 the calculation is done for the parametric
evolution (dashed line given by Eq.(\ref{e13})),
while in Fig.7 the calculation is done
with Eq.(\ref{e12}) for finite $V$ values.
The latter figure should be compared with the
upper panel of Fig.4. The agreement
is very good unless the velocity $V$ is small.
See later discussion of the QM-adiabatic regime.
The spreading $\delta E$ at the end of the pulse is
better illustrated in Fig.8.

There are three possible strategies for evaluating
the bandprofile. The first is to use the semiclassical
strategy with Eq.(\ref{eA3}). The second is to make a careful
numerical evaluation of the matrix elements, and to
use the definition Eq.(\ref{eA1}). This gives the thin
line in Fig.~5. However, the most practical and
economical procedure is simply to deduce the bandprofile
from the spreading profile $P(r)$ that correspond
to the smallest $\delta x$ value. This is the thick
line in Fig.~5, which we regard as the most appropriate
estimate.

We found out that the LRT formula Eq.(\ref{e12})
is {\em not} sensitive to the way in which the
bandprofile is evaluated,
unless the $|n-m|=1$ elements of the $\mbf{F}_{nm}$
matrix dominate the result. Let us denote by $\sigma$
the root-mean-square magnitude of these matrix elements.
The sensitivity to $\sigma$ happens in the $V$ regime
which is determined by the
condition $V < (\Delta)^2/(\hbar\sigma)$,
which is the adiabaticity condition.
This sensitivity can be used as a practical
tool for the determination of the adiabatic regime.
In the upper panel of Fig.~8 we display the result
of the LRT calculation upon setting $\sigma=0$.
We see that the LRT calculation
implies QM-adiabatic behavior for $V<3$.

If $V$ is well away from the adiabatic regime,
then near-neighbor level transitions have negligible
contribution. In such case the LRT calculation is not
sensitive to the exact value of $\sigma$.
As $V$ becomes smaller, there is a larger relative
weight to the near-neighbor matrix elements.

Deep in the QM-adiabatic regime the mean-level energy
difference is resolved much {\em before} the levels are mixed,
and therefore, as a result of recurrences, the
probability stays concentrated in the initial level.
This is of course a leading order description.
In fact we cannot neglect higher order corrections.

In the adiabatic regime the spreading is dominated
by transition between neighboring levels (whereas outside
of the adiabatic regime the contribution of neighboring level
transitions can be neglected).  Therefore, it is only in
the adiabatic regime where the quantum-mechanical
calculation (using Eq.(\ref{e12})) gives results that are
different from the classical expectation \cite{ophir}.
Is it possible to witness a regime where
the quantum mechanical (rather than the classical)
LRT prediction is observed? Apparently the answer
is positive, but not in a typical numerical experiment.
The reason is that almost always the Landau-Zener
(non LRT) mechanism for energy spreading takes over \cite{wilk}.

We can verify the dominance of Landau-Zener
mechanism as follows. The theoretical prediction
is $\delta E(t) \propto \sqrt{D_E t}$ with
$D_E \propto V^{3/2}$. In the upper panel of Fig.~8
we plot the spreading at the end of half pulse
period ($t=T/2=A/V$) and at the end of full
pulse period ($t=T=2A/V$). Hence we expect
$\delta E \propto V^{1/4}$.
The agreement with this expectation is quite good
if we consider the spreading after half pulse period.
At the end of full pulse period we get that
the spreading is larger than expected. This is
apparently due to the non-adiabatic nature of the
$V\mapsto -V$ switching.

\section{Discussion and Conclusions}

The numerical study that we have presented
should be regarded as an application of
a general procedure for the analysis of energy spreading.
In the summary below we re-order the
stages in this analysis in the way which
is implied by the results of the previous sections.

The {\bf first step} in the analysis is
to find the band profile. This can be done
using the semiclassical recipe (Appendix B)
without any need to make heavy numerical simulations.
Then it is possible to determine $\delta E(t)$
by using the LRT formula Eq.(\ref{e12}).

The {\bf second step} in the analysis is
to determine the adiabatic $V$~regime.
This is done by checking whether
the LRT calculation of $\delta E(t)$
is sensitive to $\sigma$.
In order to get in the adiabatic regime
LRT-based quantum corrections
to the classical result,
we have to take the level spacing
statistics into account \cite{ophir,robbins}.
We have pointed out the difficulty in observing
such corrections. Rather we can
improve over the LRT prediction by
taking into account either higher
order or Landau-Zener corrections
to perturbation theory.

The {\bf third step} in the analysis is to calculate
the parametric perturbative profile $P^{\tbox{prt}}(r)$.
This is associated with getting
a rough {\em estimate} for the core width $\Gamma$.
Also the parametric scales $\delta x_c$ and
$\delta x_{\tbox{prt}}$ can be determined
on the basis of this analysis.
The former is determined from inspecting~$\Gamma$,
while the latter is determined by
comparing $\delta E$ of the LRT calculation
to the $P^{\tbox{prt}}$~based prediction.

The {\bf fourth step} in the analysis is
to distinguish  between the perturbative
and the non-perturbative regime.
For this purpose we have to look for the timing
of the departure from parametric behavior,
as in the lower panel of Fig.4.
For $\delta E$ one can use the
LRT based calculation of Eq.(\ref{e12}).

The {\bf fifth step} in the analysis is
to identify the sudden regime. For this purpose we have
to eliminate the "sudden time" from Eq.(\ref{e11}).
The problem is to estimate $\Gamma(t)$
in the non-perturbative regime. If we have
only the bandprofile as an input, then
we can use the rough estimate $\Gamma(t)\sim\delta E(t)$.
This is based on the assumption that in
the non-perturbative regime the energy distribution
is characterized by a single energy scale.

The {\bf sixth step} in the analysis is
to make a simulation of the parametric evolution.
This is a relatively heavy task, but it is
still much easier than making finite $V$ simulation.
[It requires merely diagonalizations, while
a temporal simulation requires an iterative procedure].
The main non-trivial effects that we have found
in the analysis of the parametric evolution were:
\begin{minipage}{\hsize}
\vspace*{0.2cm}
\begin{itemize}
\setlength{\itemsep}{0cm}
\item Higher order tails grow up.
\item A non-perturbative core region develops.
\item The core-to-tail transitions are suppressed.
\item The tails are "pushed out".
\item Tail components are swallowed by the expanding core.
\end{itemize}
\vspace*{0.0cm}
\end{minipage}
The last two items are in the spirit of
the "core-tail" theory, but go beyond
the perturbative approximation Eq.(\ref{e8}).
Knowledge of the parametric evolution allows
accurate determination of $\Gamma$, leading
to a refined determination of the $V$~regimes.

The {\bf seventh step} in the analysis is
to make simulation of the actual evolution.
We would like to emphasize that this is the first time
that the different dynamical scenarios (corresponding
to the different $V$ regimes) are illustrated in a numerical
simulation. The only other numerical studies
(mainly \cite{diego,rsp}) were too restricted in scope,
and did not contain analysis of the stages in
the evolution of the energy {\em distribution}.

The only non-perturbative effect on the response
that we have discussed so far is the Landau-Zener
correction to the spreading in the adiabatic regime.
Are there any other non-perturbative effects
that affect the response?
The immediate tendency is to regard LRT as the
outcome of standard first order perturbation
theory (Appendix~D). Then the question that arises is
what happens if Eq.(\ref{eD5}) does not apply?

Let us recall the answer in case of
the parametric evolution of $P(n|m)$.
As $\delta x$ becomes larger we should be
worried regarding the implications of
having the effects that are listed at
the end of Section~9.
Having expanding "core" that "pushes out" the
tails, and having growing higher-order
tail components, may suggest that the
first order calculation of the variance (Eq.(\ref{eD7}))
should be "corrected", and should include
"higher order" terms. Does it mean
that Eq.(\ref{e13}) underestimate the spreading?
Or maybe we should assume that Eq.(\ref{e8})
provides the correct "trend" of higher order corrections?
The tails in the vicinity
of the core are growing slower, which can be
regarded as a suppression of core-to-tail
transitions due to the mixing. Consequently
we would conclude that Eq.(\ref{e13})
is an overestimation of spreading.
Moreover, if we take Eq.(\ref{e8}) too seriously,
beyond its regime of validity, we would conclude
that the spreading has saturation for
large $\delta x$.

For the parametric evolution these possible
speculations turn out to be {\em wrong}.
The above effects are exactly balanced, and the LRT
formula Eq.(\ref{e13}) remains exact beyond any order
of perturbation theory, which means that it
is exact even in the non-perturbative regime
where perturbation theory is not applicable.
This claim has a simple derivation \cite{lds}.
Note also that if the system has a classical limit,
then the validity of Eq.(\ref{e13}) can be established
deep in the non-perturbative regime, where
the semiclassical approximation becomes reliable.

The question is whether this delicate balance
is violated in case of finite $V$.
For example, it may be that due to incomplete core-tail
recurrences we shall have enhanced spreading
(compared with LRT).
Unlike the parametric case, we do not have
a theoretical proof that excludes such
a possibility. In fact the contrary is the case.
We have demonstrated that for an artificial
(random matrix theory) model,
the LRT formula {\em cannot be trusted}
in the non-perturbative regime.
Whether such effect is possible also
for "quantized" system that possess
a good classical limit has been left
an open question.

The possibility of having deviation from LRT
in the non-perturbative regime was one important
motivation for the present research.
Clearly we did not witness such effect
in our simulations (Fig.8). This reflects
that there is a clash between the semiclassical
limit and the RMT limit.

\appendix

\section{\newline The spectral function~$\tilde{F}_t(\omega)$}

Given a function $f(t')$ that describes the shape
of the driving pulse during the time interval $0<t'<t$,
we define a spectral function $\tilde{F}_t(\omega)$
by Eq.(\ref{e_3}).
This spectral function describes the
spectral content of the driving pulse.
Below we list some useful driving schemes.
We use the notation $\Theta()$, which is defined
by $\Theta(\mbox{\small False})=0$
and $\Theta(\mbox{\small True})=1$.

For step function we have
\begin{eqnarray} \label{eB1}
f(t') &=& \Theta(0<t') \\
\tilde{F}_t(\omega) &=& 1
\end{eqnarray}
For rectangular pulse we have
\begin{eqnarray} \label{eB2}
f(t') &=& \Theta(0<t'<t) \\
\tilde{F}_t(\omega) &=& |1-\mbox{e}^{i\omega t}|^2
\end{eqnarray}
If it is followed by a negative pulse pulse
we get
\begin{eqnarray} \label{eB3}
f(t') &=& \Theta(0<t'<t)
-\Theta( {\small \frac{T}{2}}<t'<t) \\
\tilde{F}_t(\omega) &=& |1-2\mbox{e}^{i\omega \frac{T}{2}}
+\mbox{e}^{i\omega t}|^2
\ \ \mbox{for} \ t {>} {\small \frac{T}{2}}
\end{eqnarray}
For linear driving we have
\begin{eqnarray} \label{eB4}
f(t') &=& t' \\
\tilde{F}_t(\omega) &=& t^2 (\sinc(\half\omega t))^2
\end{eqnarray}
where $\sinc(\cdot)=\sin(\cdot)/(\cdot)$.
Note that for very large $t$ we have
\begin{eqnarray} \label{eB5}
\tilde{F}_t(\omega) \sim t \times 2\pi\delta(\omega)
\end{eqnarray}
Finally, for triangular pulse
\begin{eqnarray} \label{eB6}
f(t') &=& 2\frac{t'}{T}\Theta\left(0<t'<\frac{T}{2}\right) +
\nonumber \\
\ & \ & 2(1{-}\frac{t'}{T})\Theta\left(\frac{T}{2}<t'<T\right)
\end{eqnarray}
we get
\begin{eqnarray} \label{eB7}
\tilde{F}_t(\omega) &=&
\left[\frac{2}{T}\right]^2
t^2 (\sinc(\half\omega t))^2)
\ \ \mbox{for} \ 0{<}t{<}{\small \frac{T}{2}}
\nonumber \\
\tilde{F}_t(\omega) &=&
\left[\frac{2}{T}\right]^2
\left|\frac{1-2\mbox{e}^{i\omega \frac{T}{2}}+\mbox{e}^{i\omega t}}
{\omega}\right|^2
\ \ \mbox{for} \ {\small \frac{T}{2}}{<}t{<}T
\nonumber
\end{eqnarray}

\section{\newline The spectral function~$\tilde{C}(\omega)$}

Let us denote by $\mbf{F}_{mn}$ the matrix
representation of some quantized observable $F(Q,P)$,
in the basis which is determined by some
quantized chaotic Hamiltonian ${\cal H}(Q,P)$.
The bandprofile of $\mbf{F}_{mn}$ is conveniently
characterized by the spectral function
\begin{eqnarray} \label{eA1}
\tilde{C}(\omega) \ = \
\sum_{n(\ne m)}
\left|\mbf{F}_{nm}\right|^2
\ 2\pi\delta\left(\omega-\frac{E_n{-}E_m}{\hbar}\right)
\end{eqnarray}
with implicit average over the reference state $m$.
This is the power spectrum of the fluctuating
quantity ${\cal F}(t)$, whose classical definition
is ${\cal F}(t)=F(Q(t),P(t))$, with a corresponding
quantum-mechanical definition within the Heisenberg picture.
The power spectrum of a fluctuating quantity ${\cal F}(t)$
is defined as the Fourier transform of the corresponding
correlation function $C(\tau)$.

Chaotic systems are characterized by fast decay
of dynamical correlations. We assume a separation
of time scales between the (short) classical
correlation time, and the (long) quantum-mechanical
Heisenberg time.  Thus, for times during which we can
ignore the recurrences, we expect the following
quantal-classical correspondence:
\begin{eqnarray} \label{eA2}
C(\tau) \ \ \approx \ \ C^{\tbox{cl}}(\tau)
\end{eqnarray}
This correspondence implies that the envelope
of $\tilde{C}(\omega)$ is given
by $\tilde{C}^{\tbox{cl}}(\omega)$,
and the following semiclassical expression
for the matrix elements follows \cite{mario}:
\begin{eqnarray} \label{eA3}
\left\langle\left|\mbf{F}_{nm}
\right|^2\right\rangle
\ \ \approx \ \
\frac{\Delta}{2\pi\hbar} \
\tilde{C}^{\tbox{cl}}\left(\frac{E_n{-}E_m}{\hbar}\right)
\end{eqnarray}
Taking into account the level spacing statistics,
we deduce the following relation \cite{ophir}:
\begin{eqnarray} \label{eA4}
\tilde{C}(\omega) \ \ \approx \ \
\hat{R}(\omega) \
\tilde{C}^{\tbox{cl}}(\omega)
\end{eqnarray}
where $\hat{R}(\omega) \propto
\langle \sum_n \delta(\omega-\omega_{nm})\rangle_m$
is the two-point correlation function of the
energy spectrum (Fourier transform of the spectral
form factor). For the Gaussian orthogonal ensemble (GOE)
of random matrix theory (RMT) the following result
is well known:
\begin{eqnarray} \label{eA5}
\left. \hat{R}(\omega) \right|_{\tbox{GOE}} \ = \
1-(\sinc(\pi\hbar\omega/\Delta))^2
\end{eqnarray}
and more generally it is common to
assume $\hat{R}(\omega) \sim \omega^{\beta}$
for small $\omega$.

\section{Classical derivation of the LRT formula}

Consider ${\cal H}(Q,P;x(t))$,
and define the following time dependent
quantities:
\begin{eqnarray} \label{eC1}
{\cal F}(t) &=& -\frac{\partial{\cal H}}{\partial x} (Q(t),P(t);x(t))  \\
{\cal E}(t) &=& {\cal H}(Q(t),P(t);x(t)) \\
{\cal E}'(t) &=& {\cal H}(Q(t),P(t);x(0))
\end{eqnarray}
Note that ${\cal E}(t)$ is the energy in the conventional
sense, while ${\cal E}'(t)$ is the energy using a "fixed basis".
We have the following relations
\begin{eqnarray} \label{eC2}
\frac{d {\cal E}(t)}{dt} \ = \ \ \ \
\frac{\partial {\cal H}}{\partial t} \ \ \ \ &= \
-\dot{x}(t) {\cal F}(t)
\\ \label{eC3}
\frac{d {\cal E}'(t)}{dt} \ = \
- [{\cal H},{\cal H}_0] \ &\approx \
\delta x(t) \dot{{\cal F}}(t)
\end{eqnarray}
The latter approximated equality strictly holds
if the perturbation ${\cal H}-{\cal H}_0$ is linear
with respect to the perturbation parameter
$\delta x = x - x_0$.
From the equations above it follows that
\begin{eqnarray} \label{eC4}
{\cal E}(t)-{\cal E}(0) \ &=& \
- \int_0^t \dot{x}(t') {\cal F}(t') dt'
\\ \label{eC5}
{\cal E}'(t)-{\cal E}'(0) \ & \approx & \
\int_0^t \delta x(t') \dot{\cal F}(t') dt'
\end{eqnarray}
Obviously the two latter expressions coincide
for a cycle ($x(t)=x(0)$). Squaring Eq.(\ref{eC4}),
and performing a microcanonical average
over the (implicit) initial conditions $(Q(0),P(0))$
one obtains
\begin{eqnarray} \label{eC6}
\delta E(t)^2 \ =  \ A^2
\int_0^t \int_0^t \dot{f}(t') \dot{f}(t'') C(t'-t'') dt' dt''
\end{eqnarray}
This can be written as Eq.(\ref{e12}).

\section {First order perturbation theory and LRT}

For convenience we assume that the perturbation
is linear in $\delta x = x - x_0$. In complete analogy
with the classical analysis we can work either
using "fixed basis", or else we can use the
proper $x$-dependent basis (the so called
adiabatic representation). The respective
matrix representations of the Hamiltonian are
\begin{eqnarray} \label{eD1}
{\cal H} \mapsto \mbf{E} + \delta x(t) \mbf{F}
\\ \label{eD2}
{\cal H} \mapsto \mbf{E} + \dot{x}(t) \mbf{W}
\end{eqnarray}
where $\mbf{E}$ is a diagonal matrix.
Note that in the first equation $\mbf{E}$
and $\mbf{F}$ are calculated for $x=x_0$,
while in the second equation there is
an implicit $x(t)$ dependence.  
The matrix elements of $\mbf{F}$ are
\begin{eqnarray} \label{eD3}
\mbf{F}_{nm} \ = \
\Big\langle n
\Big| \frac{\partial{\cal H}}{\partial x}
\Big| m
\Big\rangle
\end{eqnarray}
The off diagonal matrix elements
of $\mbf{W}$ are
\begin{eqnarray} \label{eD4}
\mbf{W}_{nm} \ = \
\frac{\hbar}{i}
\Big\langle n \Big| \frac{d}{dt} m \Big\rangle \ = \
i\frac{\hbar}{E_n{-}E_m}
\mbf{F}_{nm}
\end{eqnarray}
and we use the `gauge' convention
$\mbf{W}_{nm}=0$ for $n{=}m$.
(Only one parameter is being changed and
therefore Berry's phase is not an issue).
The derivation of Eq.(\ref{eD2}) is 
standard, and can be found in Section~11 
of \cite{frc}.

Using first order perturbation theory
with Eq.(\ref{eD2}) we get
\begin{eqnarray} \label{eD5}
P_t(n|m) = \delta_{nm} +
A^2 \tilde{F}_t\left( \frac{E_n{-}E_m}{\hbar} \right)
\times \left| \frac{\mbf{W}_{nm}}{\hbar} \right|^2
\end{eqnarray}
Note that for a cycle ($f(t)=f(0)=0$),
the same result is obtained via
first order perturbation theory with Eq.(\ref{eD1}).
As a global approximation Eq.(\ref{eD5}) is valid
only in the standard perturbative regime. In the extended
perturbative regime it is valid only for
the first order tail region (see Section~8).
The expression for the first order tail region can
be written in a concise way as
\begin{eqnarray} \label{eD6}
P_t(n|m) =
A^2 \tilde{F}_t(\omega_{nm})
\times
\frac{\Delta}{2\pi\hbar} \tilde{C}(\omega_{nm})
\
\left[\frac{1}{\hbar\omega_{nm}}\right]^2
\end{eqnarray}
The corresponding global approximation
is given by Eq.(\ref{e8}).

Assuming that the variance is dominated by the
first order tail component, we get the LRT result
\begin{eqnarray} \label{eD7}
\delta E(t)^2
\ &=&  \ \sum_{n} P_t(n|m) \times (E_n-E_m)^2
\nonumber \\
\ &=&  \ A^2 \times
\int_{-\infty}^{\infty}
\frac{d\omega}{2\pi\hbar}
\tilde{F}_t(\omega)
\tilde{C}(\omega)
\end{eqnarray}
An implicit average over $m$ is assumed.
Note that the latter formula does
not contain $\hbar$. This "restricted"
quantal-classical correspondence holds only
for the variance. Higher moments of the
energy distribution are typically much smaller
compared with the classical expectation,
and scale like $\hbar$ to the power of the
moment order minus~2.


\ \\

\noindent
{\bf Acknowledgments:}
It is our pleasure to thank Tsampikos Kottos and
Eduardo Vergini for useful discussions.
The research was supported in part
by the Israel Science Foundation (grant No.11/02).
DAW gratefully acknowledges support from CONICET
(Argentina). Research grants by CONICET and ECOS-SeTCIP.

\ \\

\newpage



\clearpage

\noindent
\epsfig{figure=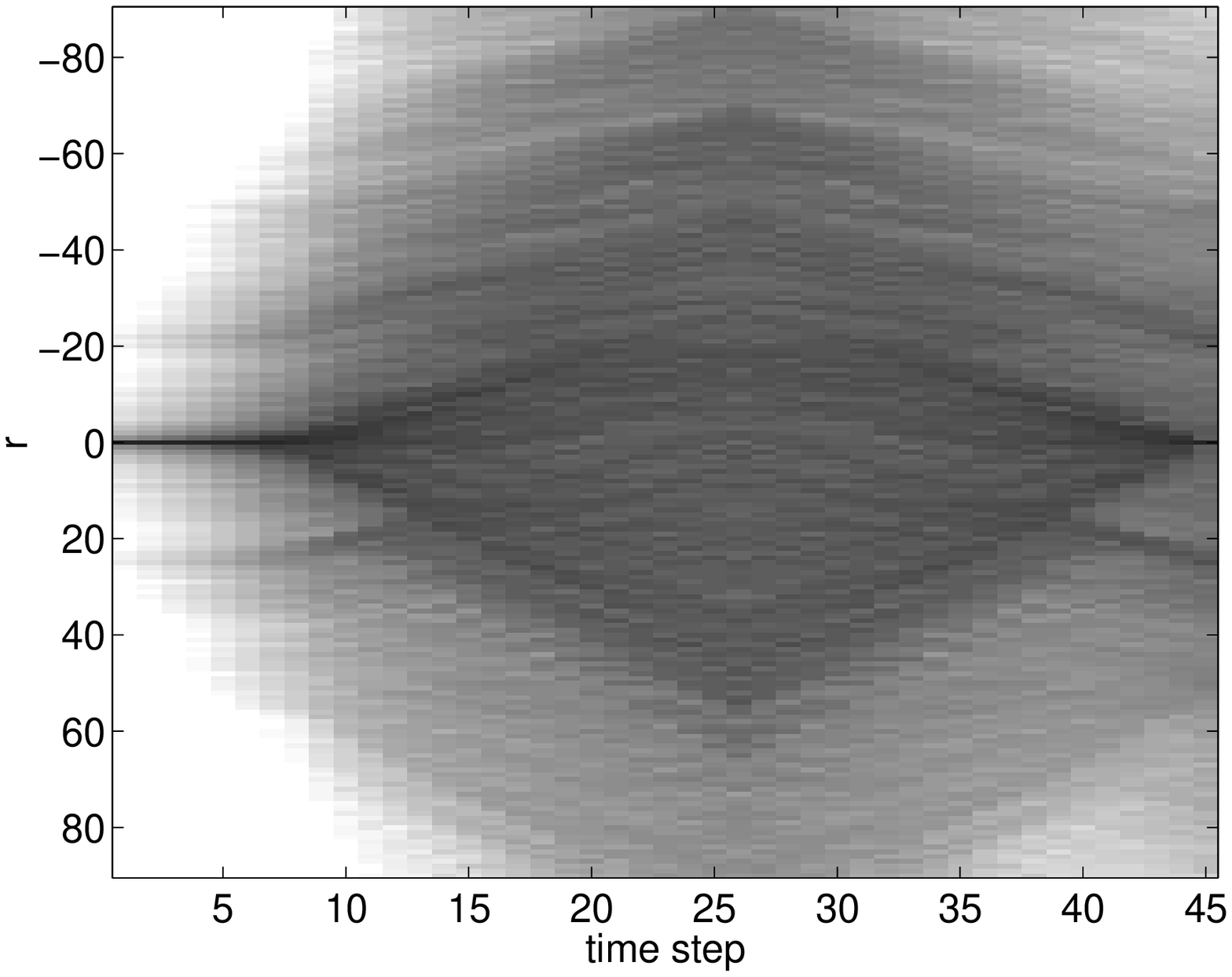,width=\hsize} \\
\epsfig{figure=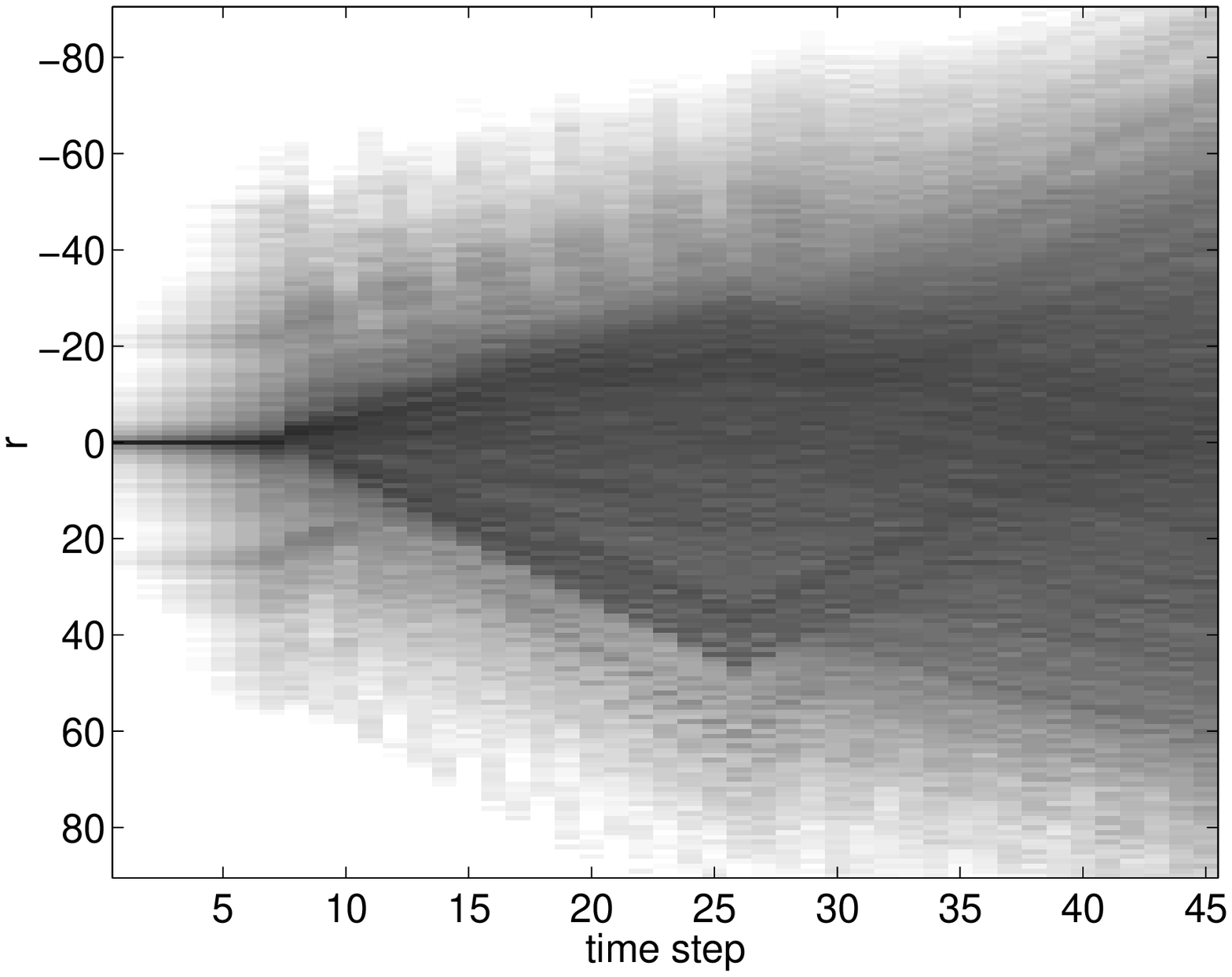,width=\hsize} \\
{\footnotesize
Fig.1. Images of time evolution within one-pulse period
for $V=100$  (upper panel) and for $V=1$ (lower panel).
Each column is a profile of the probability distribution $P_t(r)$
for a different time step $t$. The first~7 time steps are
log spaced, while the rest are linearly spaced.
The $V=100$ evolution is predominantly  parametric.}

\newpage

\noindent
\epsfig{figure=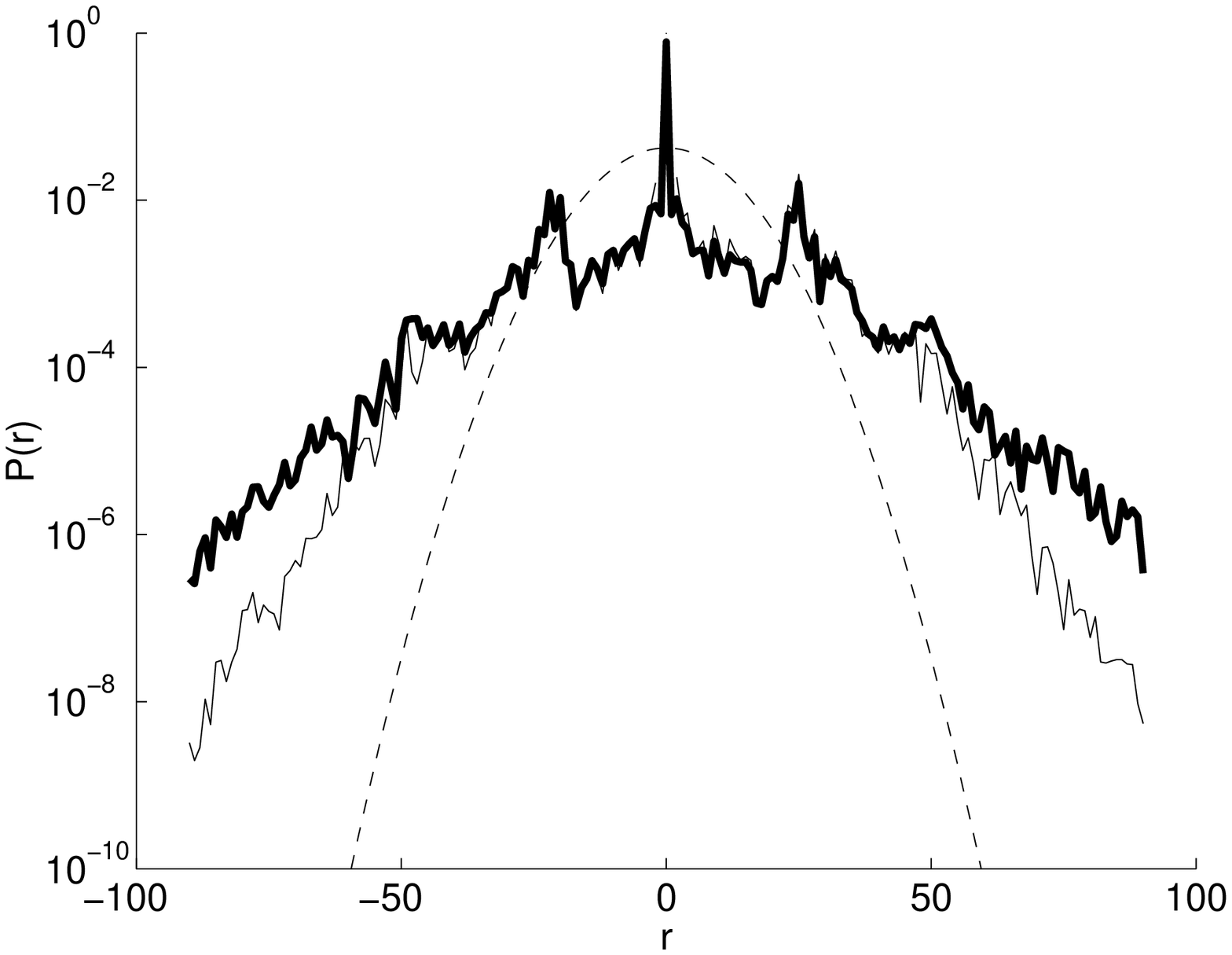,width=\hsize} \\
\epsfig{figure=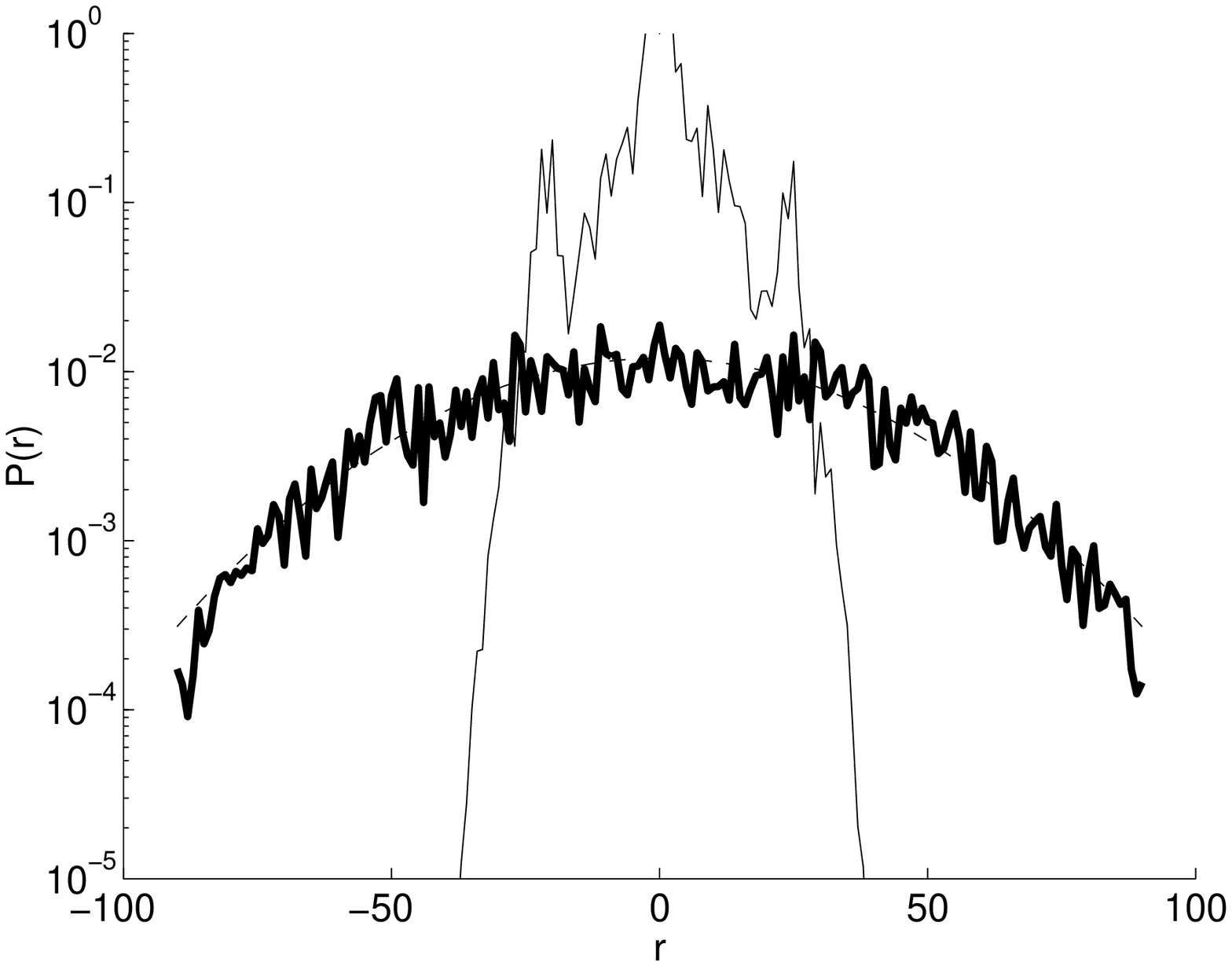,width=\hsize} \\
{\footnotesize
Fig.2. Spreading profile at the end of one pulse period
for $V=100$ (upper panel) and for $V=10$ (lower panel).
The thin solid lines are from standard
first order perturbation theory, which means
$P_t^{\tbox{prt}}$ of Eq.(\ref{e8}) with $\Gamma=0$.
The dashed lines are for Gaussian
with the same variance.}


\clearpage

\noindent
\epsfig{figure=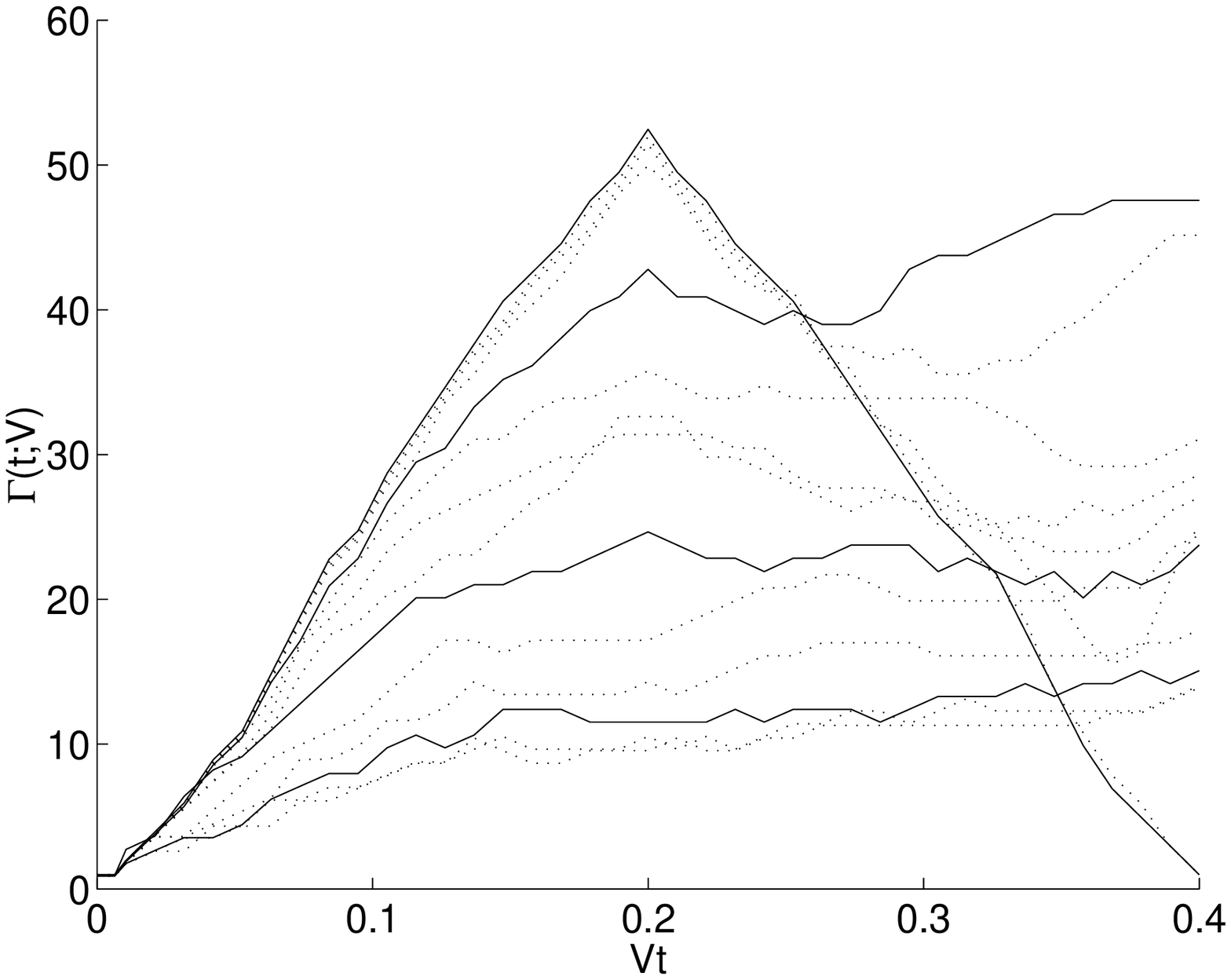,width=\hsize} \\
\epsfig{figure=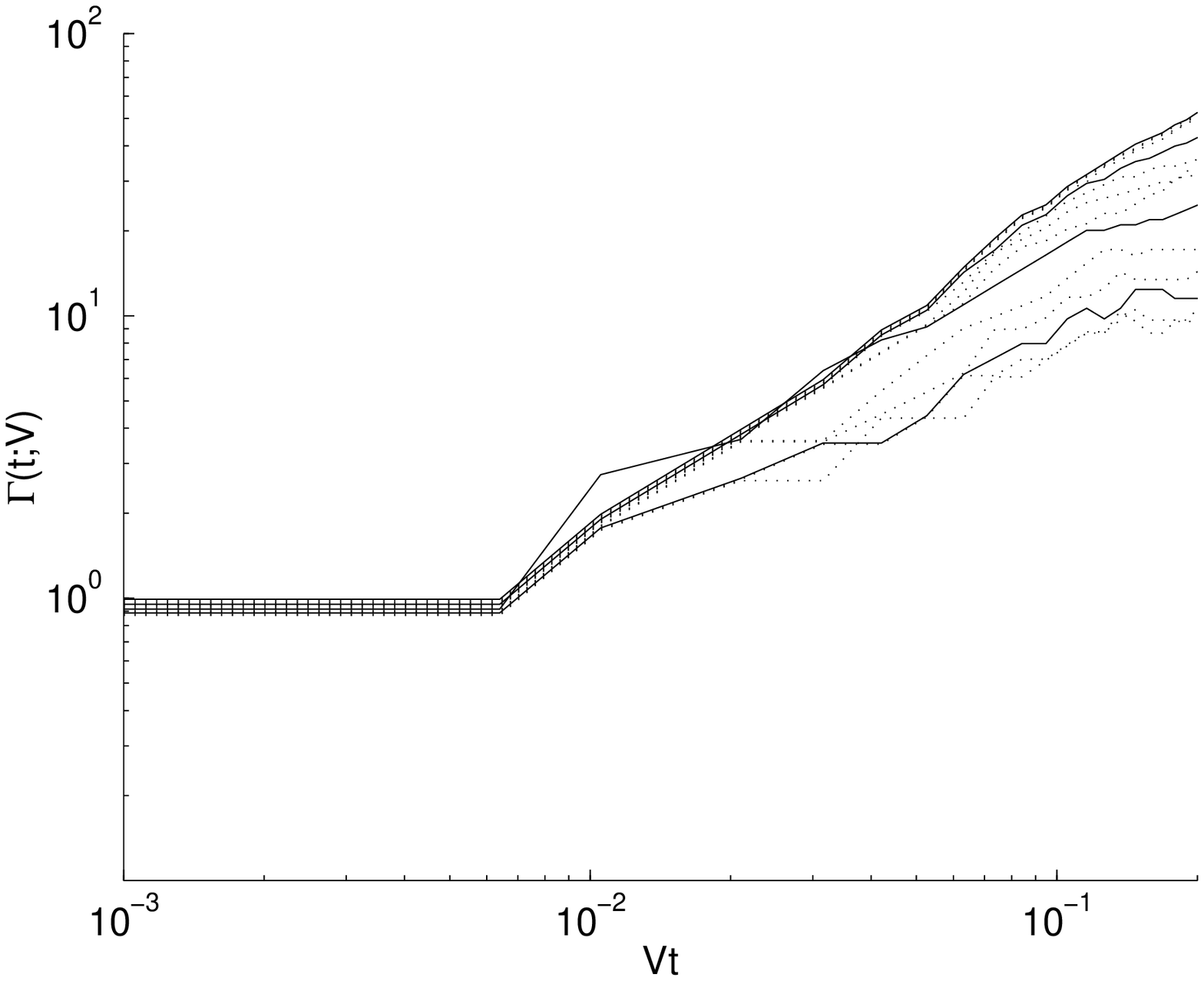,width=\hsize} \\
{\footnotesize
Fig.3. The width $\Gamma(t)$ as a function of time.
The numerical determination of $\Gamma$ is explained
after Eq.(\ref{e6}). The horizontal axis is
the scaled time $Vt$.
The different curves correspond to the different velocities
$V = 100, 80, 40, 20, 10, 7, 5, 3, 1, 0.5, 0.2, 0.1, 0.07, 0.05$.
The solid lines highlight the velocities
$V=100$ (upper most), and $V=10$, and $V=1$, and $V=0.1$.
The lower panel is the same as in the upper figure,
but in log-log scale, and only half period is displayed.}

\newpage

\noindent
\epsfig{figure=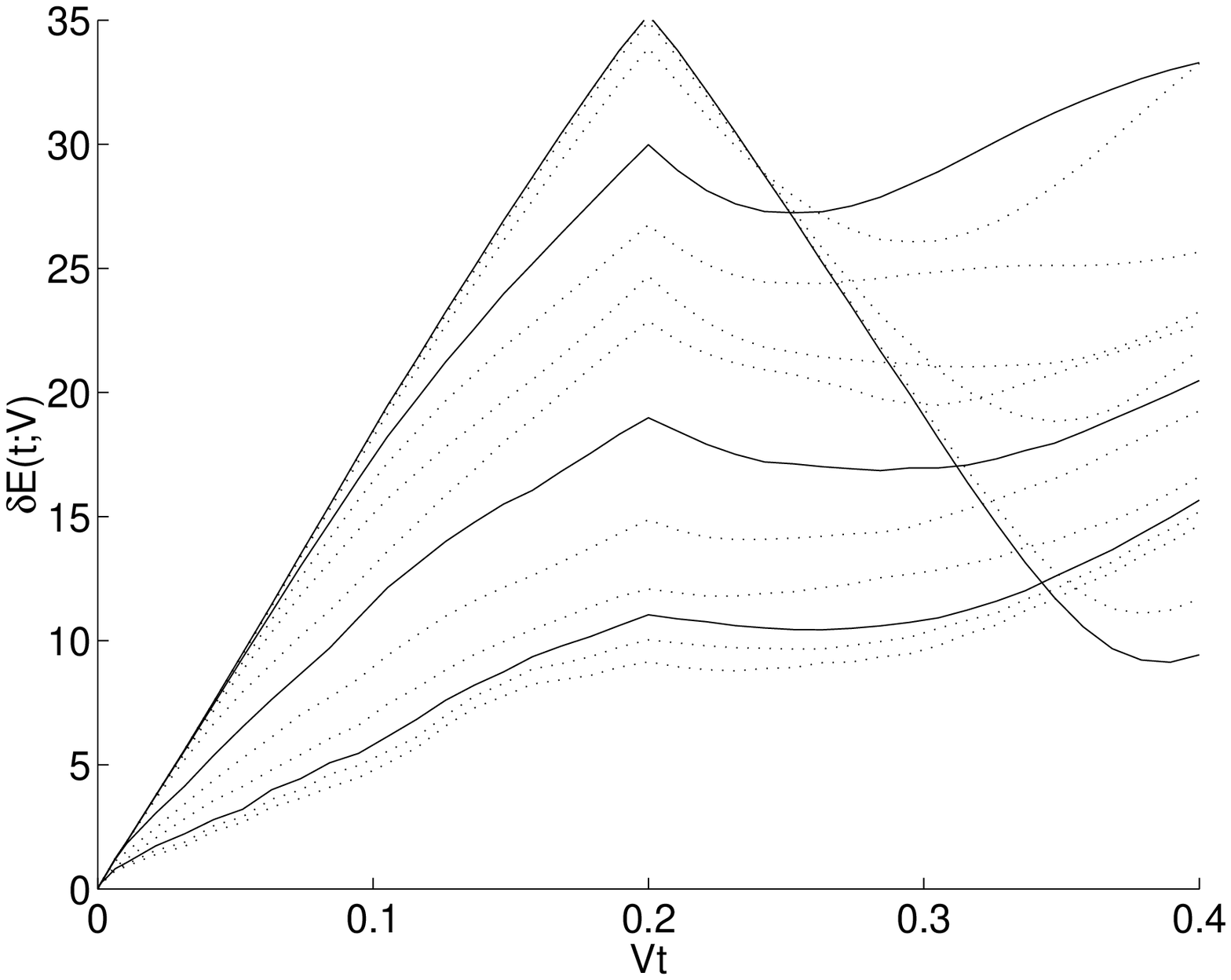,width=\hsize} \\
\epsfig{figure=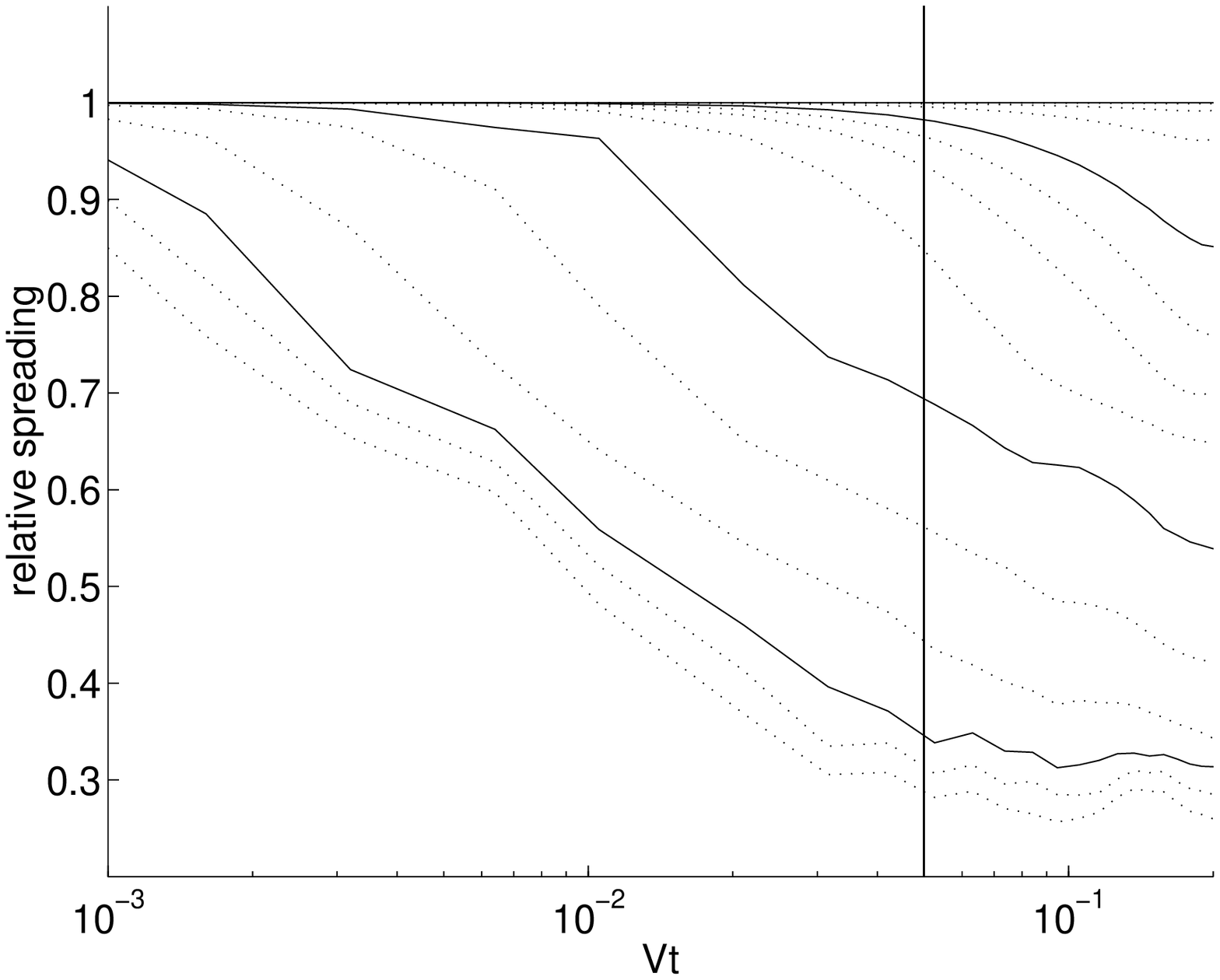,width=\hsize} \\
{\footnotesize
Fig.4. The spreading $\delta E(t)$ of Eq.(\ref{e7})
as a function of time, for different $V$ simulations.
The upper panel is in one to one correspondence
with the upper panel of the previous figure.
In the lower panel the data of the upper
panel are presented in a different way:
the vertical axis is the relative spreading;
The horizontal axis is log scale;
And only half period is displayed.
The location of $\delta x_{\tbox{prt}}=0.05$
is indicated by vertical line.
See the text for more details.}


\newpage

\noindent
\epsfig{figure=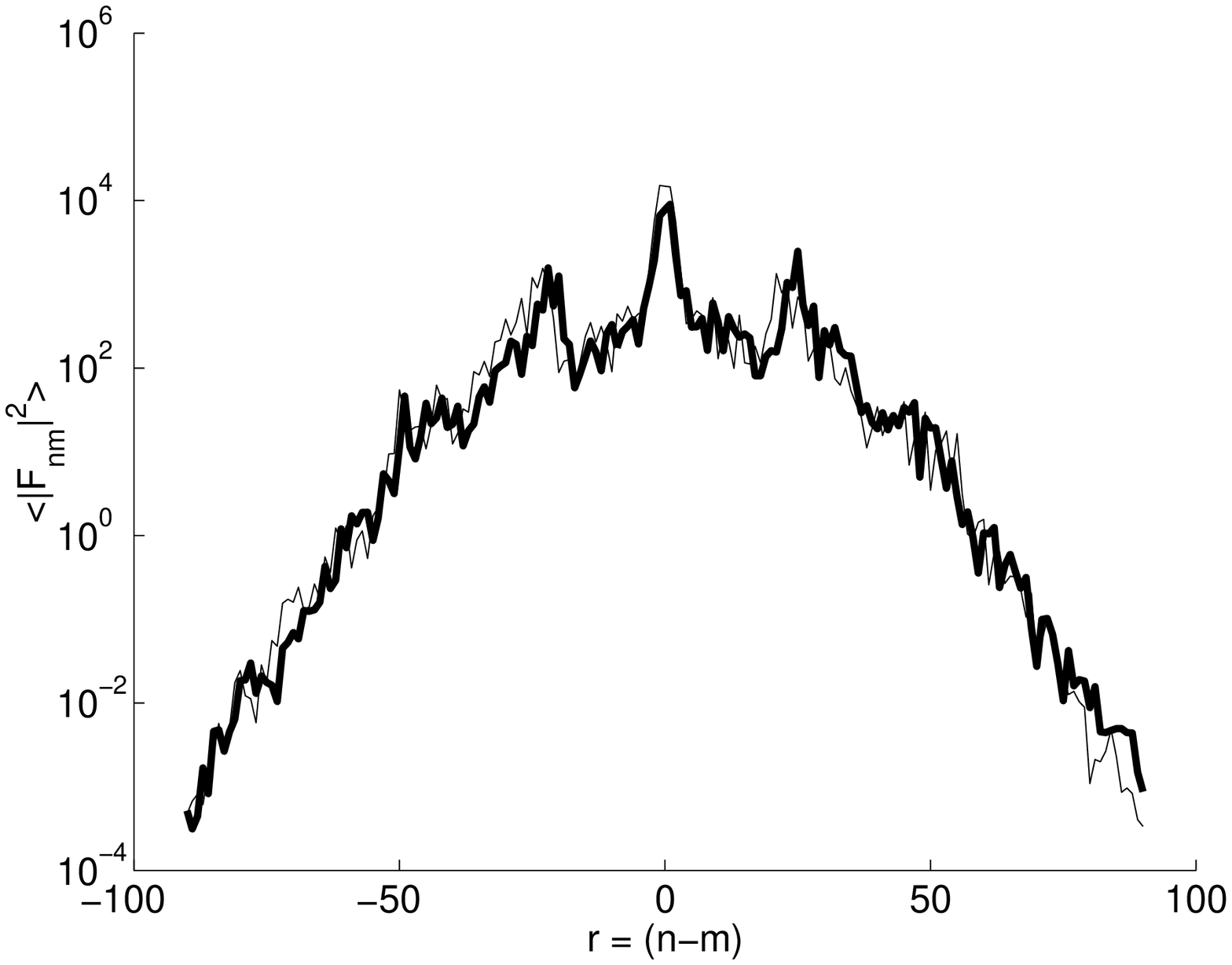,width=0.9\hsize} \\
{\footnotesize
Fig.5. The band profile, as defined in Appendix~B.
The two curves are the outcome of two
different numerical procedures:
The thin line is based on direct
evaluation of matrix elements,
while the thick line is deduced
form the evolution over an
infinitesimal time step
(see Section~12 for further details).
A third possible procedure (not displayed)
is to use the semiclassical recipe Eq.(\ref{eA3})

\ \\

\noindent
\epsfig{figure=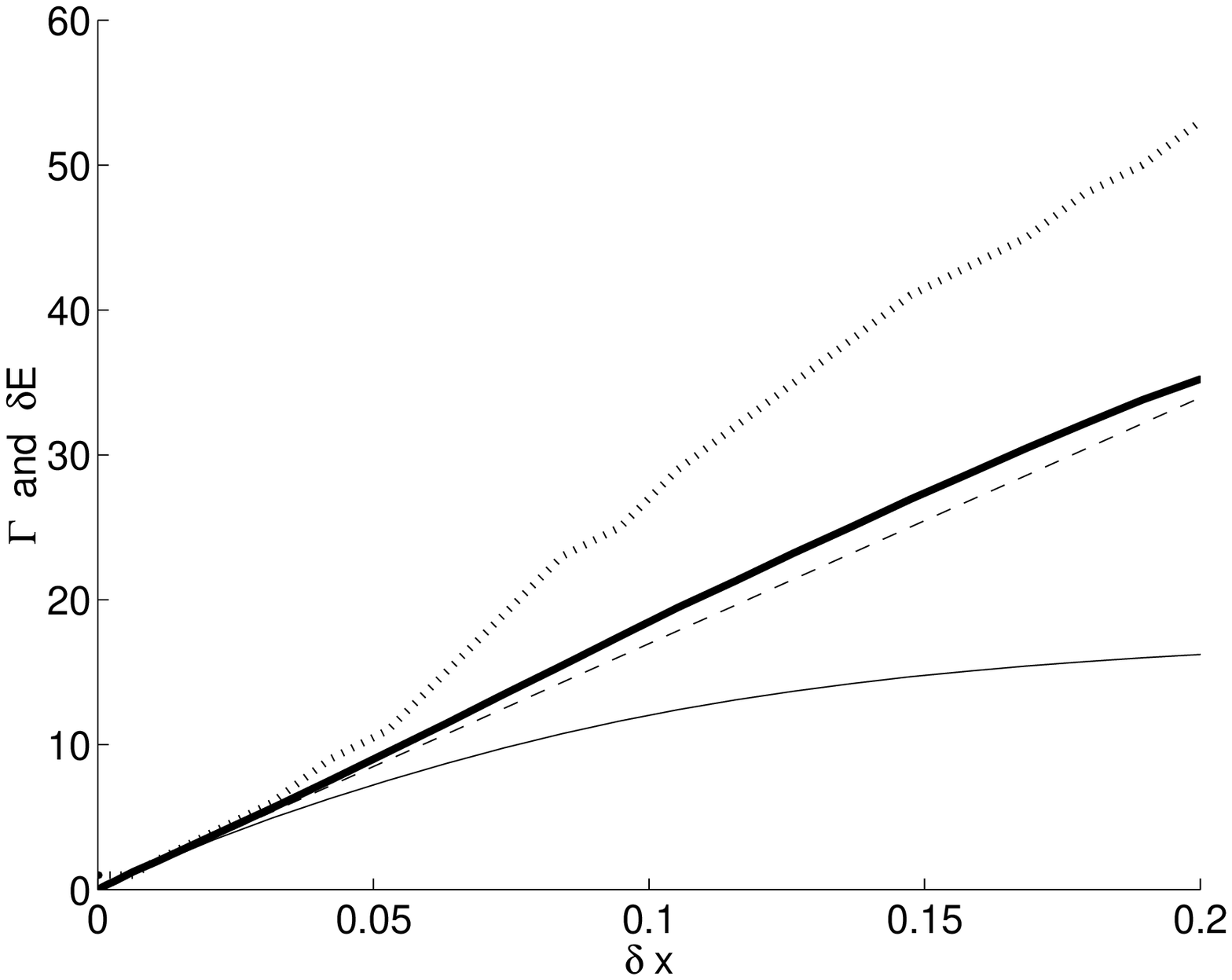,width=0.9\hsize} \\
\epsfig{figure=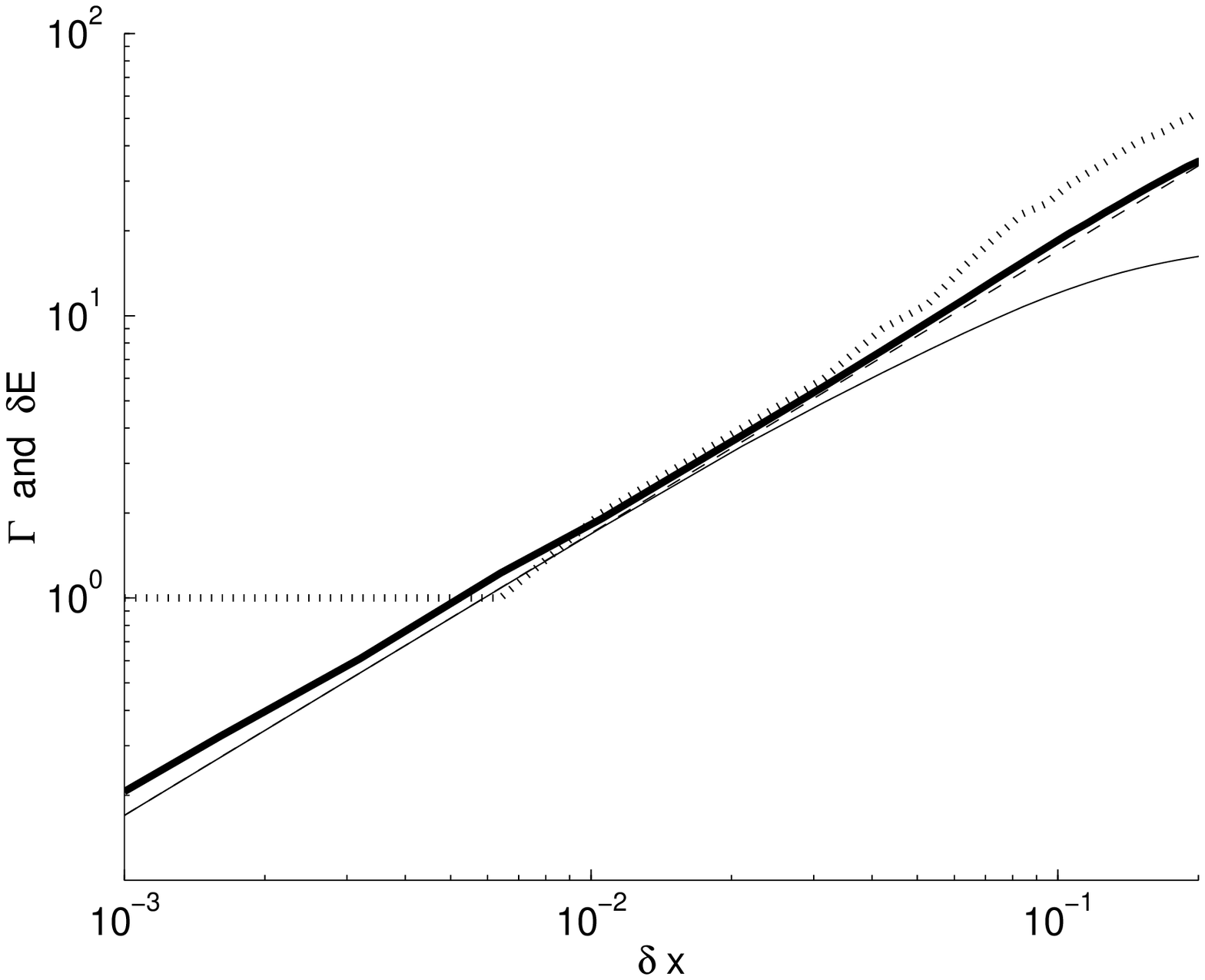,width=0.9\hsize} \\
{\footnotesize
Fig.6. Parametric evolution:
The horizontal axis is $\delta x$
either in linear scale (upper panel)
or in logarithmic scale (lower panel).
The dotted line is the width $\Gamma$
from (\ref{e6}).
The thick line is $\delta E$ from (\ref{e7}),
while the dashed line is the LRT
estimate Eq.(\ref{e13}).
The thin line is the perturbative estimate
Eq.(\ref{e7}) using Eq.(\ref{e8}).
From this plots we can determine
$\delta x_c=0.006$ and $\delta x_{\tbox{prt}}=0.05$. }


\newpage

\noindent
\epsfig{figure=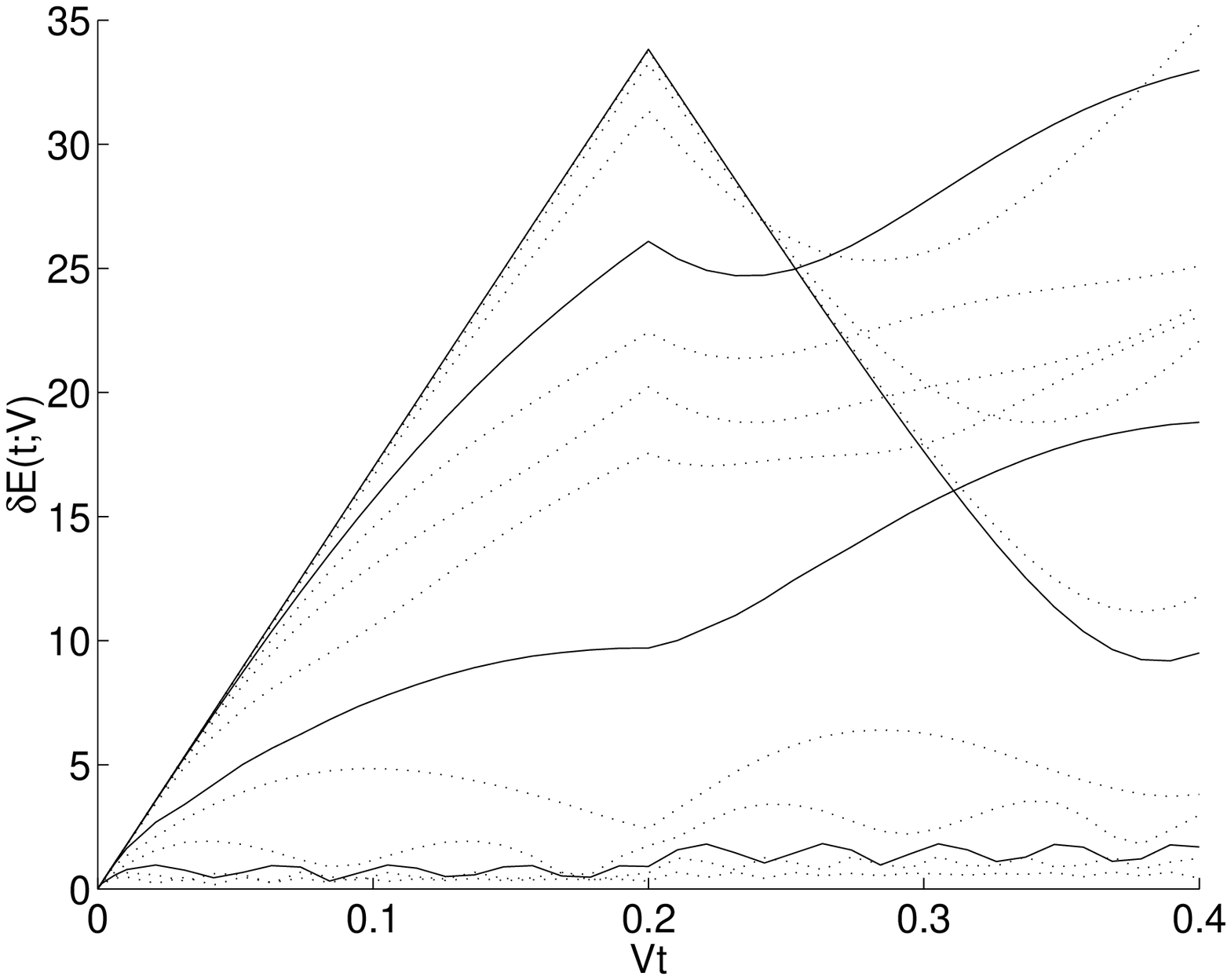,width=\hsize} \\
{\footnotesize
Fig.7. LRT calculation of the spreading versus time,
using Eq.(\ref{e12}) with the bandprofile
as an input. The horizontal axis is $Vt$.
This Figure is in one to one correspondence
with the upper panel of Fig.4. One observes
that for small~$V$ the calculation underestimates
the observed spreading. See discussion in Section~12.
}

\ \\

\ \\

\noindent
\epsfig{figure=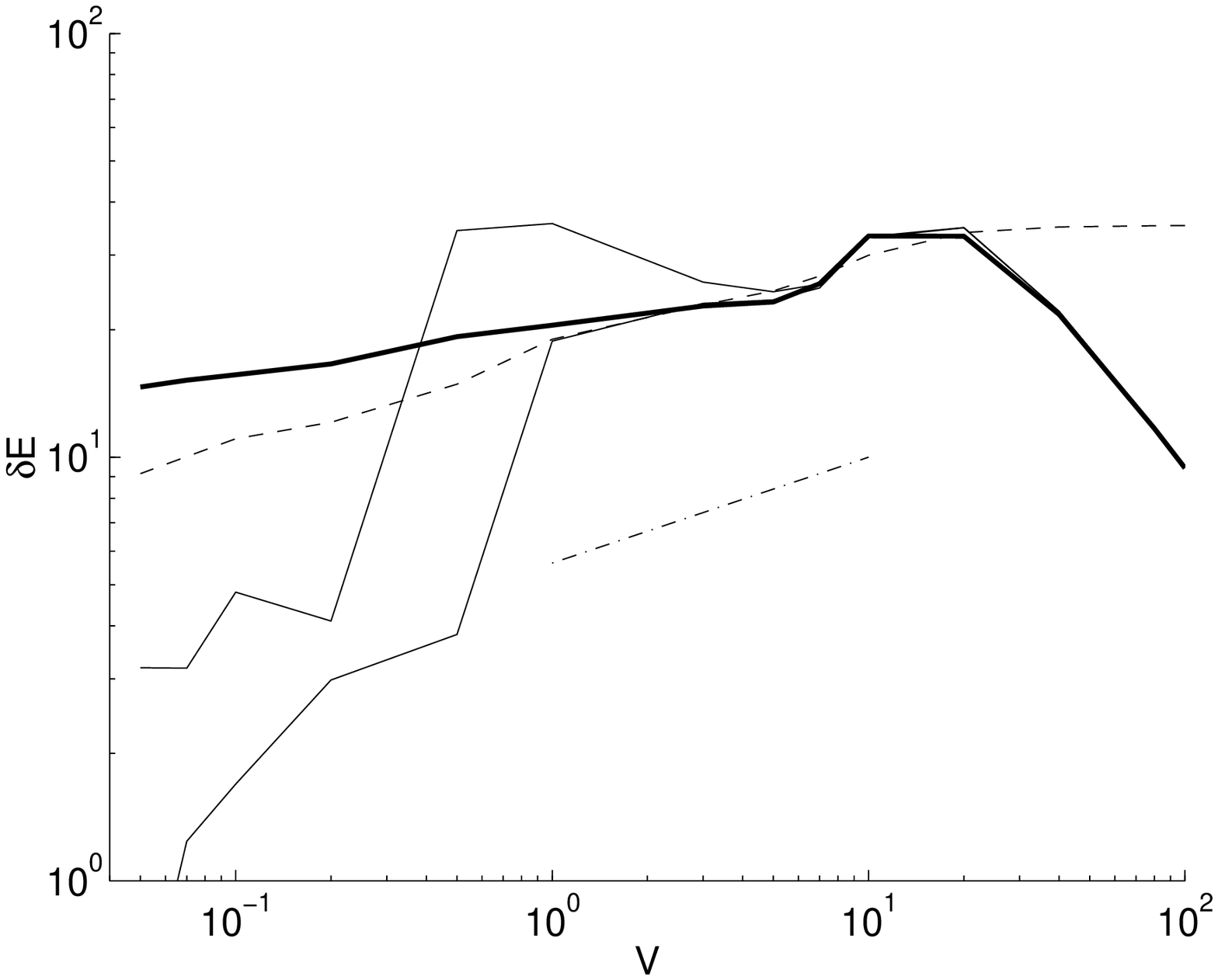,width=\hsize} \\
\epsfig{figure=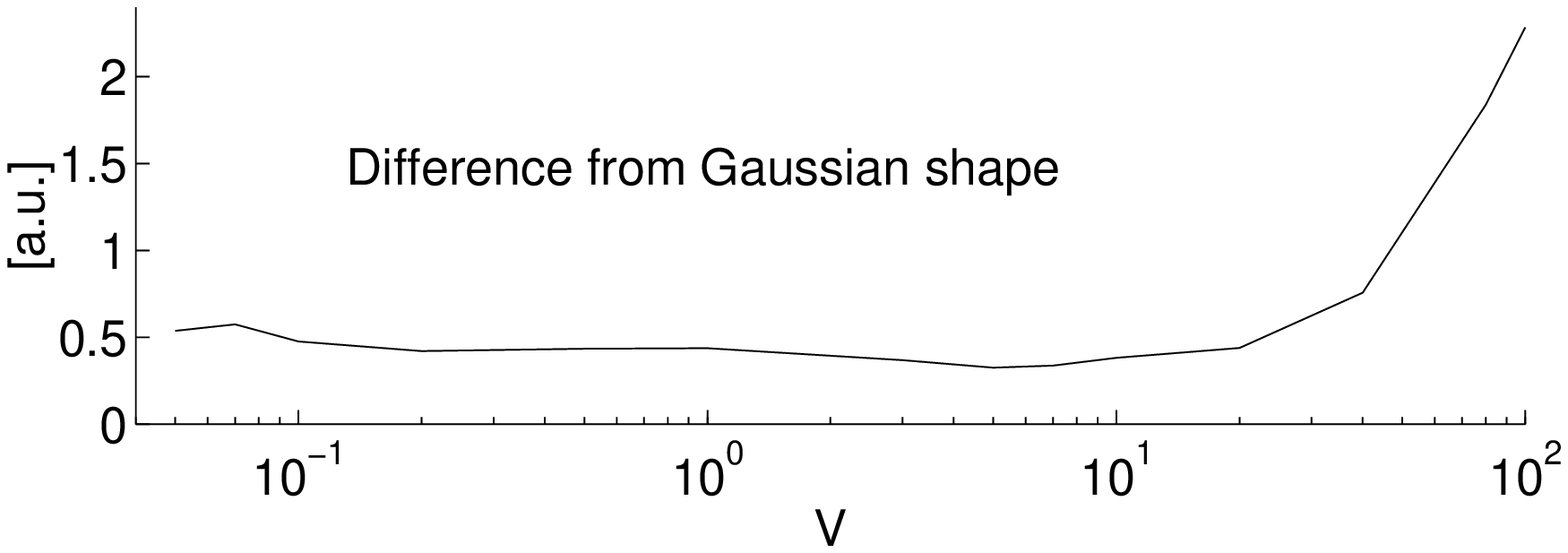,width=\hsize}
{\footnotesize
Fig.8. upper panel: Response for one-pulse versus $V$.
The Thick line is $\delta E$ at the
end of the pulse. The thin (solid and dotted) lines
correspond to the LRT calculation Eq.(\ref{e12}).
For the calculation of the lower thin lines
we have set the near-level coupling $\sigma=0$.
The dashed line is the spreading at the
end of half pulse period. The dash-dot line
is the slope that corresponds to Landau-Zener
spreading mechanism. The lower panel is the
difference Eq.(\ref{e_diff}) from Gaussian line shape.}


\begin{thebibliography}{99}



\bibitem{lecnotes}
For a pedagogical presentation, including references
see \cite{dsp} and \cite{vrn}, which can be
downloaded from http://www.bgu.ac.il/$\sim$dcohen.


\bibitem{dsp}
D. Cohen, "Driven chaotic mesoscopic systems,dissipation and decoherence",
in "Dynamical Semigroups: Dissipation, Chaos, Quanta",
{\em Proceedings of the 38 Winter School of Theoretical Physics},
Ed. P. Garbaczewski and R. Olkiewicz (Springer-Verlag, in press).


\bibitem{vrn}
D. Cohen in "New directions in quantum chaos",
{\em Proceedings of the International School
of Physics "Enrico Fermi", Course CXLIII},
Edited by G. Casati, I. Guarneri and U. Smilansky,
(IOS Press, Amsterdam 2000).





\bibitem{infwell_a}
S.W. Doescher and M.H. Rice,
Am. J. Phys. {\bf 37}, 1246 (1969).

\bibitem{infwell_b}
A.J. Makowski and S.T. Dembinski,
Physics Letters A {\bf 154}, 217 (1991).

\bibitem{jose} 
J.V. Jose and R. Cordery,
Phys. Rev. Lett. {\bf 56}, 290 (1986).




\bibitem{qkr} 
For review and references see \cite{qkr1} and \cite{qkr2}.

\bibitem{qkr1}
S. Fishman in "Quantum Chaos",
{\em Proceedings of the International School
of Physics "Enrico Fermi", Course CXIX},
Ed. G. Casati, I. Guarneri and U. Smilansky
(North Holland 1991).


\bibitem{qkr2}
M. Raizen in "New directions in quantum chaos",
{\em Proceedings of the International School
of Physics "Enrico Fermi", Course CXLIII},
Edited by G. Casati, I. Guarneri and U. Smilansky
(IOS Press, Amsterdam 2000).





\bibitem{crs}
D. Cohen, Phys. Rev. Lett. {\bf 82}, 4951 (1999).

\bibitem{rsp}
D. Cohen and T. Kottos, Phys. Rev. Lett. {\bf 85}, 4839 (2000).

\bibitem{frc}
D. Cohen, Annals of Physics {\bf 283}, 175 (2000).




\bibitem{VS}
E. Vergini and M. Saraceno, Phys. Rev. E, {\bf 52}, 2204 (1995).

\bibitem{vergini_thesis}
E. Vergini, PhD thesis (Universidad de Buenos Aires, 1995).

\bibitem{barnett_thesis}
A. Barnett, PhD thesis (Harvard University, 2000).

\bibitem{diego}
D.A. Wisniacki and E. Vergini,
Phys. Rev. E {\bf 59}, 6579 (1999).

\bibitem{prm}
D. Cohen, A. Barnett and E.J. Heller,
Phys. Rev. E {\bf 63}, 46207 (2001).






\bibitem{ophir}
O.M. Auslaender and S. Fishman,
Phys. Rev. Lett. {\bf 84}, 1886 (2000);
J. Phys. A {\bf 33}, 1957 (2000).

\bibitem{robbins}
J.M. Robbins and M.V. Berry,
J. Phys. A {\bf 25}, L961 (1992). \\
M.V. Berry and J.M. Robbins,
Proc. R. Soc. Lond. A {\bf 442}, 659 (1993). \\
M.V. Berry and E.C. Sinclair,
J. Phys. A {\bf 30}, 2853 (1997).

\bibitem{wilk}
M. Wilkinson, J. Phys. A {\bf 21}, 4021 (1988);
J. Phys. A {\bf 20}, 2415 (1987).



\bibitem{wlf}
A. Barnett, D. Cohen and E.J. Heller, J. Phys. A {\bf 34}, 413 (2001);
Phys. Rev. Lett. {\bf 85},  1412 (2000).

\bibitem{wld}
D. Cohen, Phys. Rev. E {\bf 65}, 026218 (2002).

\bibitem{lds}
D. Cohen and T. Kottos,  Phys. Rev. E {\bf 63}, 36203 (2001).



\bibitem{mario}
M. Feingold and A. Peres, Phys. Rev. A {\bf 34} 591, (1986).
M. Feingold, D. Leitner, M. Wilkinson, Phys. Rev. Lett. {\bf 66}, 986 (1991);
M. Wilkinson, M. Feingold, D. Leitner, J. Phys. A {\bf 24}, 175 (1991);
M. Feingold, A. Gioletta, F. M. Izrailev, L. Molinari,
Phys. Rev. Lett. {\bf 70}, 2936 (1993).


\end{thebibliography}
\end{document}